\documentclass[longauth]{aa}
\usepackage{graphicx}
\usepackage{txfonts}
\usepackage[colorlinks=true,citecolor=blue,linkcolor=black]{hyperref}
\usepackage{siunitx}
\usepackage{xcolor}
\usepackage{float}
\usepackage{bm}
\usepackage{soul,xcolor}
\setstcolor{red}
\definecolor{alizarin}{rgb}{0.82, 0.1, 0.26}

\newcommand{\oiii}{\textup{[O\,\textsc{iii}]}}
\newcommand{\nii}{\textup{[N\,\textsc{ii}]}}
\newcommand{\sii}{\textup{[S\,\textsc{ii}]}}
\newcommand{\oi}{\textup{[O\,\textsc{i}]}}

\newcommand{\hii}{\textup{H}\,\textsc{ii}}
\newcommand{\ha}{\textup{H}\ensuremath{\alpha}}
\newcommand{\hb}{\textup{H}\ensuremath{\beta}}

\newcommand{\um}{$\rm \mu m$}

\begin{document}

   \title{Calibrating mid-infrared emission as a tracer of obscured star formation on H\textsc{ii}-region scales in the era of JWST}
    
\author{Francesco~Belfiore\inst{\ref{arcetri}}\fnmsep\thanks{\email{francesco.belfiore@inaf.it}}
\and
Adam~K.~Leroy\inst{\ref{ohio}}
\and
Thomas~G.~Williams \inst{\ref{oxford}}
\and
Ashley~T.~Barnes\inst{\ref{eso}}
\and
Frank Bigiel\inst{\ref{UBonn}}
\and
Médéric~Boquien\inst{\ref{uta}}
\and
Yixian~Cao \inst{\ref{mpe}}
\and
J\'er\'emy Chastenet \inst{\ref{UGent}}
\and
Enrico~Congiu\inst{\ref{eso_chile}}
\and
Daniel~A.~Dale\inst{\ref{wyo}} 
\and
Oleg~V.~Egorov\inst{\ref{rechen}}
\and 
Cosima~Eibensteiner \inst{\ref{UBonn}}
\and 
Eric~Emsellem \inst{\ref{eso}, \ref{cral}}
\and
Simon~C.~O.~Glover\inst{\ref{zah}}
\and   
Brent~Groves\inst{\ref{uwa}}  
\and
Hamid~Hassani \inst{\ref{ualberta}}
\and
Ralf~S.~Klessen\inst{\ref{zah},\ref{zw}}
\and
Kathryn~Kreckel\inst{\ref{rechen}}
\and
Lukas~Neumann\inst{\ref{UBonn}}
\and
Justus~Neumann\inst{\ref{mpia}}
\and
Miguel~Querejeta\inst{\ref{oan}}
\and
Erik~Rosolowsky \inst{\ref{ualberta}}
\and
Patricia~Sanchez-Blazquez \inst{\ref{ucm},\ref{iparcos}}
\and
Karin Sandstrom \inst{\ref{ucsd}}
\and
Eva~Schinnerer\inst{\ref{mpia}}
\and
Jiayi~Sun\inst{\ref{McMaster},\ref{CITA}}
\and
Jessica~Sutter\inst{\ref{ucsd}}
\and
Elizabeth~J.~Watkins\inst{\ref{rechen}}
 }

\institute{INAF — Osservatorio Astrofisico di Arcetri, Largo E. Fermi 5, I-50125, Florence, Italy\label{arcetri}
   \email{francesco.belfiore@inaf.it}
    \and Department of Astronomy, The Ohio State University, 140 West 18th Avenue, Columbus, OH 43210, USA\label{ohio}
    \and Sub-department of Astrophysics, Department of Physics, University of Oxford, Keble Road, Oxford OX1 3RH, UK\label{oxford}
    \and  European Southern Observatory, Karl-Schwarzschild Stra{\ss}e 2, D-85748 Garching bei M\"{u}nchen, Germany \label{eso}
    \and Argelander-Institut f\"ur Astronomie, Universit\"at Bonn, Auf dem H\"ugel 71, D-53121 Bonn, Germany\label{UBonn}
     \and Instituto de Alta Investigación, Universidad de Tarapacá, Casilla 7D, Arica, Chile\label{uta}
    \and Max-Planck-Institut f\"ur extraterrestrische Physik (MPE), Giessenbachstrasse 1, D-85748 Garching, Germany\label{mpe}
     \and Sterrenkundig Observatorium, Ghent University, Krijgslaan 281-S9, 9000 Gent, Belgium\label{UGent}
    \and European Southern Observatory (ESO), Alonso de C\'ordova 3107, Casilla 19, Santiago 19001, Chile \label{eso_chile}
    \and Department of Physics and Astronomy, University of Wyoming, Laramie, WY 82071, USA\label{wyo}
    \and Astronomisches Rechen-Institut, Zentrum f\"ur Astronomie der Universit\"at Heidelberg, M\"onchhofstra{\ss}e 12-14, D-69120 Heidelberg, Germany\label{rechen}
     \and Univ Lyon, Univ Lyon1, ENS de Lyon, CNRS, Centre de Recherche Astrophysique de Lyon UMR5574, F-69230 Saint-Genis-Laval France\label{cral}
    \and Universit\"at Heidelberg, Zentrum f\"ur Astronomie, Institut f\"ur theoretische Astrophysik, Albert-Ueberle-Stra{\ss}e 2, D-69120, Heidelberg, Germany\label{zah}
    \and  International Centre for Radio Astronomy Research, University of Western Australia, 7 Fairway, Crawley, 6009, WA, Australia\label{uwa}
    \and Universit\"at Heidelberg, Interdisziplin\"ares Zentrum f\"ur Wissenschaftliches Rechnen, Im Neuenheimer Feld 205, D-69120 Heidelberg, Germany\label{zw}
    \and Max-Planck-Institute for Astronomy, K\"onigstuhl 17, D-69117 Heidelberg, Germany\label{mpia}
    \and Observatorio Astronómico Nacional (IGN), C/Alfonso XII, 3, E-28014 Madrid, Spain\label{oan}   
    \and Department of Physics, University of Alberta, Edmonton, Alberta, T6G 2E1, Canada\label{ualberta}
    \and Departamento de F\'isica de la Tierra y Astrof\'isica, Universidad Complutense de Madrid, E-28040 Madrid, Spain \label{ucm}
    \and Instituto de F\'{\i}sica de Part\'{\i}culas y del Cosmos IPARCOS, Facultad de CC Físicas, UCM, E-28040, Madrid, Spain \label{iparcos}
    \and Department of Astronomy \& Astrophysics, University of California, San Diego, 9500 Gilman Drive, San Diego, CA 92093, USA \label{ucsd}
    \and Department of Physics and Astronomy, McMaster University, 1280 Main St. West, Hamilton ON L8S 4M1 Canada\label{McMaster}
    \and Canadian Institute for Theoretical Astrophysics (CITA), University of Toronto, 60 St George Street, Toronto, ON M5S 3H8, Canada\label{CITA}
}

\date{Received XXX; accepted XXX}

 
\abstract{
Measurements of the star formation activity on cloud scales are fundamental to uncovering the physics of the molecular cloud, star formation, and stellar feedback cycle in galaxies. Infrared (IR) emission from small dust grains and polycyclic aromatic hydrocarbons (PAHs) are widely used to trace the obscured component of star formation. However, the relation between these emission features and dust attenuation is complicated by the combined effects of dust heating from old stellar populations and an uncertain dust geometry with respect to heating sources. We use images obtained with NIRCam and MIRI as part of the PHANGS--JWST survey to calibrate the IR emission at 21\um, and the emission in the PAH-tracing bands at 3.3, 7.7, 10, and 11.3~\um\ as tracers of obscured star formation. We analyse  $\sim$ 20~000 optically selected \hii\ regions across 19 nearby star-forming galaxies, and benchmark their IR emission against dust attenuation measured from the Balmer decrement. We model the extinction-corrected \ha\ flux as the sum of the observed \ha\ emission and a term proportional to the IR emission, with $a_{IR}$ as the proportionality coefficient.
A constant $a_{IR}$ leads to an extinction-corrected \ha\ estimate which agrees with those obtained with the Balmer decrement with a scatter of $\sim$ 0.1 dex for all bands considered. Among these bands, 21~\um\ emission is demonstrated to be the best tracer of dust attenuation. The PAH-tracing bands underestimate the correction for bright \hii\ regions, since in these environments the ratio of PAH-tracing bands to 21~\um\ decreases, signalling destruction of the PAH molecules. For fainter \hii\ regions all bands suffer from an increasing contamination from the diffuse infrared background. We present calibrations that take this effect into account by adding an explicit dependence on 2~\um\ emission or stellar mass surface density.
}
\titlerunning{MIR emission tracing obscured star formation with JWST}
\authorrunning{F. Belfiore}
\maketitle


\section{Introduction} \label{sec:intro}

Interstellar dust biases our view of star formation, obscuring the ultraviolet (UV) and optical emission of newly formed massive stars. The energy absorbed by dust at short wavelengths is re-radiated in the IR, making observations of dust emission a powerful probe of embedded star formation. Multi-wavelength studies, at both low and high redshift, have highlighted the importance of the IR to provide a full accounting of star formation rates (SFRs) in galaxies \citep{Calzetti2007, Wuyts2011, Kennicutt2012, Gruppioni2013, Madau2014, Calzetti2020}. 

The thermal dust emission spectrum peaks in the far-IR. In the limit where the dust heating is dominated by young stellar populations, the far-IR emission can be directly related to the embedded SFR using energy balance \citep{Kennicutt1998}. However, such a bolometric measure of dust luminosity is sensitive to heating by older stellar populations \citep{Cortese2008}. This heating term contributes progressively more to the cooler, longer wavelength far-IR emission, leading to diffuse `IR cirrus'. On the other hand, the mid-IR (MIR) region around 24~$\mu $m, corresponding to the IRAS 25\um, the \textit{Spitzer} MIPS 24 $\mu $m band, or the WISE W4 (22 $\mu $m) filter, traces emission of small, hot grains. While not directly relatable to the bolometric dust luminosity without further assumptions on the shape of the dust SED, emission at 24 \um\ has the advantage of tracing a hotter dust component, and therefore being less sensitive to IR cirrus.

Several authors have advocated the use of 24~\um\ fluxes, in combination with either UV or \ha\ recombination line emission, to account for both the attenuated and un-attenuated component of the SFR \citep{Hirashita2003, Kennicutt2007, Calzetti2007, Kennicutt2009, Wuyts2011a, Hao2011, Leroy2012, Kennicutt2012, Jarrett2013, Catalan-Torrecilla2015, Lesie2018}. These hybrid recipes have been demonstrated to be successful for both integrated galaxies and kpc-scale sub-galactic regions in local galaxies \citep{Calzetti2007, Leroy2012, Belfiore2023}. However, dust heating from old stellar populations can still constitute a significant fraction of a galaxy's integrated emission at 24~\um\ \citep{Bendo2012, Groves2012, Crocker2013, Lu2015, Boquien2016, Viaene2017} and complicates the association of the MIR with star formation in observations that have low spatial resolution \citep{Leroy2012}. 

In addition to 24\um, several authors have studied the potential of using the flux in PAH features or the brightness of PAH-tracing bands (most commonly the IRAC 8~\um\ band) to probe obscured star formation \citep{Peeters2004, Shipley2016a, Maragkoudakis2018, Whitcomb2020}.
These calibrations offer significant advantages with respect to the longer mid-IR wavelengths in terms of angular resolution and sensitivity \citep{Calzetti2020, Li2020}. Moreover, PAH features remain accessible and are readily identifiable with \textit{Spitzer}, and now JWST, out to high redshifts \citep{Lutz2005, Riechers2014}. Studies of nearby galaxies at high spatial resolution have demonstrated, however, that PAH emission is suppressed relative to 24 \um\ (or total IR) emission at the position of \hii\ regions \citep{Helou2004, Bendo2006, Bolatto2007, Lebouteiller2011, Chastenet2019, Chastenet2023a, Egorov2023}. In particular, PAH emission is found to generate bright rings around \hii\ regions, probably due to PAH photo-destruction, therefore complicating the relation between PAHs and star formation \citep{Galliano2018}.

In this work, we test the use of both the MIR dust continuum at 21 \um\ and bands dominated by PAH emission to trace embedded star formation on the scale of individual \hii\ regions (${\sim} 40{-}110$ pc) in a sample of 19 nearby disc galaxies observed with the MIRI and NIRCam instruments onboard JWST as part of the PHANGS (Physics at High Angular resolution in Nearby Galaxies) JWST program \citep{Lee2023}. We compare the attenuation estimates obtained using literature calibrations based on the MIR/\ha\ ratios with matched-resolution Balmer decrement (\ha/\hb) measurements obtained from the PHANGS--MUSE integral field spectroscopy (IFS) mapping with the VLT/MUSE instrument \citep{Emsellem2022}. 

Using this JWST data we calibrate photometric measurements of the MIR continuum and PAH emission to accurately trace the obscured component of star formation on scales of individual \hii\ regions. Our analysis builds on the work of \cite{Leroy2023} and \cite{Hassani2023}, who used early data for a subset of four galaxies in our sample to demonstrate that attenuation-corrected \ha\ emission correlates well with MIR emission at 21~\um, albeit non-linearly, for regions that are bright in the MIR. 


In Sec. \ref{sec:D&M} we present our data, paying special attention to the new JWST images, and summarise the methods for inferring the dust attention.  In Sec. \ref{sec:results} we present our results concerning the use of the IR bands in correcting \ha\ for dust attenuation in \hii\ regions, and discuss the recommended recipes and potential future work in Sec. \ref{sec:discussion}. We summarise the results in Sec. \ref{sec:conclusions}.

\section{Data and Methods} 
\label{sec:D&M}

\subsection{PHANGS--JWST data}
\label{jwst_data}

We analysed 19 galaxies observed as part of the cycle 1 PHANGS--JWST Treasury program (ID 02107, PI: J. Lee). The survey consists of MIRI and NIRCam images for galaxies observed by the PHANGS--MUSE program \citep{Emsellem2022}. 
Key properties of our target galaxies are summarised in Table~\ref{tab:tab1}. 

For each target we obtained one or more NIRCam and MIRI tiles, covering roughly the same area on sky and maximising the overlap with existing MUSE observations. MIRI images were obtained in the F770W, F1000W, F1130W and F2100W filters, centred at 7.7, 10, 11.3, and 21~\um, respectively. With NIRCam we obtained images in the F200W, F300M, F335M, and F360M filters, centred at 2.0, 3.0, 3.37, and 3.6 \um, respectively. NIRCam and MIRI observations were obtained in sequence, in order to obtain a MIRI background pointing in parallel with the NIRCam observations. However, for three galaxies (IC~5332, NGC~4303, NGC~4321) all or part of the NIRCam observations were not obtained due to guide star failures. In the case of IC~5332 NIRCam data was obtained during a subsequent visit, while for NGC~4303 and NGC~4321 only one NIRCam tile was obtained out of the intended two-tile mosaic, therefore covering roughly half of the area imaged with MIRI. Because of the failure of the first NIRCam visit for IC~5332, to process its MIRI data the background frame for NGC~7496 (taken on the same day) was used. 

The observing strategy and data reduction workflow are presented in \cite{Lee2023} and Williams et al. (in preparation), respectively. Here we briefly discuss specific reduction steps of particular importance for this work, and which represent an update with respect to the early reduction process described in \cite{Lee2023}. We used the latest version of the CRDS context (jwst\_1070) and science calibration pipeline (v1.10.0). 
Secondly, we have implemented a custom background matching technique to minimise the per-pixel difference in each pair of overlapping tiles. The JWST pipeline assumes a constant background across the image, which is not the case for our observations (in many cases, one half of the image has a significantly lower `background' level than the other). Our routine avoids this by calculating the median per-pixel difference in each overlapping pair, and finding values that minimise this across the tile set. Thirdly, we have improved the absolute astrometry for the long wavelength MIRI tiles, by basing the astrometric  solution of the longer bands on that of the shortest MIRI band (F770W). As each tile in each filter is observed close in time, and using the same guide stars, this produces significantly better astrometric solutions than attempting to match each band separately. Finally, as discussed in \cite{Leroy2023} the background level of the overall image mosaics are re-scaled to match those of existing \textit{Spitzer} or WISE data. This step is necessary because we generally do not have empty sky regions that can used for background subtraction within our mosaics. 


We convolved the JWST images to match the MUSE point spread function (PSF), which is different for each target as discussed in the next subsection. The required kernels were generated using the methodology of \cite{Aniano2011} and the JWST PSFs were obtained via the WebbPSF tool \citep{Perrin2014}. Error maps produced by the JWST reduction pipeline were also convolved with the same kernels and the errors are propagated analytically through the convolution process. 
For one galaxy, NGC~4535, the MUSE PSF full width at half maximum (FWHM) (0.56$''$) is smaller than the F2100W PSF. We therefore convolve the JWST data with the smallest suitable kernel (an `aggressive' kernel in the definition of \citealt{Aniano2011}) to generate an image with a Gassian PSF. The resulting PSF FWHM is of 0.71$''$ FWHM, and the MUSE maps for this galaxy are smoothed to the same PSF.



In the NIRCam wavelength range, the PAH emission feature at 3.3 \um\ falls in the F335M medium-band filter. As discussed in \cite{Sandstrom2023}, we performed a bespoke continuum subtraction procedure using the adjoining F300M and F360M medium-band filters in order to remove stellar continuum emission and minimise the impact of the PAH emission contaminating the F360M side-band. Our procedure is based on and improves upon the scheme first suggested by \cite{Lai2020a}. In this work we refer to the resulting continuum-subtracted emission as 
\begin{equation}
\rm F335M_{\rm PAH} = F335M - F335M_{\rm cont},
\end{equation}
where $\rm F335M_{\rm cont}$ is the continuum contamination estimated using equations 10 and 11 from \cite{Sandstrom2023}. For the rest of this work we only work with the continuum-subtracted $\rm F335M_{\rm PAH}$ band.

The data for a few galaxies necessitated further by-hand masking of prominent diffraction spikes (NGC~1365, NGC~1566, NGC~7496) caused by bright AGN, evident in the F2100W images presented in Fig. \ref{fig:figAll}, further described below. 

\begin{table*}[]
	\centering
	\begin{tabular}{l c c c c c c c l}
	Galaxy	 & D & $\log(\rm M_\star)$ & $\log (\rm SFR)$ & $i$  & Copt PSF & $\theta_{\rm bm}$ &$\rm N_{HII}$  & Notes \\
	 & [Mpc] & $[\rm M_\odot]$ & $[\rm M_\odot ~yr^{-1}]$ & [deg] & [arcsec] & [pc] &   &  \\
	\hline
	IC 5332 & 9.0 & 9.67 & $-0.39$ & 26.9 &    0.87 & 38 & 545 & \\ 
    NGC 0628 & 9.8 & 10.34 & 0.24 & 8.9 &  0.92 & 44& 1712 &  grand-design spiral\\ 
    NGC 1087 & 15.9 & 9.93  & 0.12 & 42.9 & 0.92 & 71 & 912 & \\ 
    NGC 1300 & 19.0 & 10.62  & 0.07 & 31.8  & 0.89 & 82 & 944 & nuclear ring \\  
    NGC 1365 & 19.6 & 10.99 & 1.23 & 55.4   & 1.16 & 109 & 550 & AGN, nuclear ring  \\ 
    NGC 1385 & 17.2 & 9.98 & 0.32 & 44.0   & 0.77 & 64 & 913 &  \\ 
    NGC 1433 & 18.6 & 10.87 & 0.05 & 28.6  & 0.91 & 82 & 851 &  nuclear ring \\ 
    NGC 1512 & 18.8 & 10.71 & 0.11 & 42.5  & 1.25 & 114 & 477 &  nuclear ring \\ 
    NGC 1566 & 17.7 & 10.78 & 0.66 & 29.5  & 0.80 & 69 & 1561 & AGN \\ 
    NGC 1672 & 19.4 & 10.73 & 0.88 & 42.6  & 0.96 & 90 & 934 & AGN, nuclear ring\\ 
    NGC 2835 & 12.2 & 10.00 & 0.09 & 41.3  & 1.15 & 68 & 732 & \\
    NGC 3351 & 10.0 & 10.36 & 0.12 & 45.1  & 1.05 & 51 & 690 & nuclear ring\\
    NGC 3627 & 11.3 & 10.83 & 0.58 & 57.3  & 1.05 & 58 & 970 & AGN \\ 
    NGC 4254 & 13.1 & 10.42 & 0.49 & 34.4  & 0.89 & 57 & 2022 & \\
    NGC 4303 & 17.0 & 10.52 & 0.73 & 23.5  & 0.78 & 64 & 2043 &  AGN, nuclear ring, NIRCam covers half \\ 
    NGC 4321 & 15.2 & 10.75 & 0.55 & 38.5  & 1.16 & 86 & 1239 &  nuclear ring, NIRCam covers half  \\ 
    NGC 4535 & 15.8 & 10.53 & 0.33 & 44.7 & 0.71$^{\star}$ & 54 & 1097 &  AGN \\ 
    NGC 5068 & 5.2 & 9.40 & $-$0.56 & 35.7 &  1.04 & 26 & 1174 & \\ 
    NGC 7496 & 18.7 & 10.00 & 0.35 & 35.9 &   0.89 & 81  & 535 &  AGN  \\ 
    
    \hline
	\end{tabular}
	\caption{Distances are taken from the compilation of \cite{Anand2021}, stellar mass and SFR are from \cite{Leroy2021a}, inclinations from \cite{Lang2020} where available, otherwise from \cite{Leroy2021a}, and PSF FWHM of the copt MUSE data are from \cite{Emsellem2022} ($^\star$ for this galaxy the copt FWHM is smaller than the F2100W PSF, so we used an `aggressive' kernel to convolve the JWST data to a Gaussian PSF). $\theta_{\rm bm}$ gives the size of the common resolution element with no inclination correction. $\rm N_{HII}$ gives the number of HII regions from \cite{Groves2023} overlapping with the MIRI footprint, detected in F2100W, and meeting our other selection criteria (see text). The notes include information on which galaxies have partial coverage with NIRCam.}
	\label{tab:tab1}
\end{table*}

\subsection{MUSE data and the optical \hii\ region catalog}

We compare the JWST images with maps of \ha\ and \hb\ obtained with IFS from the PHANGS--MUSE program \citep{Emsellem2022}, and the associated \hii\ region masks \citep{Groves2023}. In short, individual MUSE pointings (field of view of 1$' \times 1'$) were convolved with a kernel chosen to generate a common Gaussian PSF across each galaxy mosaic which is also constant as a function of wavelength. We refer to the resulting PSF as convolved-optimised or `copt'. We adopt this copt FWHM as the target resolution for the JWST images, except in the case of NGC~4535 already noted above. 
Table~\ref{tab:tab1} summarises the adopted FWHM resolution for each galaxy. 

\hii\ region masks were derived with the \textsc{hiiphot} algorithm \citep{Thilker2000}, which separates nebulae from the surrounding emission by first defining seed regions around bright peaks, and extending them until a termination gradient in the surface brightness profile is reached. Line fluxes and key nebular properties (e.g.\ metallicity, ionisation parameter) were derived within each \hii\ region integrating the spectra inside the mask following the procedure detailed in \cite{Groves2023}. 

Fluxes in each JWST band were extracted within our \hii\ region masks by reprojecting the convolved JWST images onto the MUSE astrometric grid. Regions only partially covered by the footprint of the MIRI data or including masked or saturated pixels were discarded. Since the MIRI and NIRCam coverage in each target is not identical, and some targets lack NIRCam data coverage, some \hii\ regions have MIRI fluxes but lack NIRCam ones, or vice versa. In general, for each section of the analysis we use all \hii\ regions with valid JWST data, leading to a slightly different sample of regions when different bands are considered. We have checked that restricting the analysis to \hii\ regions that have valid fluxes in all JWST bands does not alter the results.

No local background subtraction was performed in our fiducial analysis because of the substantial crowding of sources and the resulting uncertainty intrinsic to a local background definition. However, we test the impact of this choice by performing a simple local background subtraction for each \hii\ region. We define the background as the mean surface brightness in an annulus around each region two MUSE pixels (each pixel is 0.2$''$) distant from the region boundary and three pixels in width. If such an annulus overlaps with other regions, we remove the overlap area from the background mask. This procedure is simplistic and leads to fairly large background levels. For example, at 21\um\ 25\% of the \hii\ regions in our sample have fluxes consistent with the local background with their error, and the average contrast between the flux in the \hii\ regions and in the background is 1.5. Because of the fairly arbitrary choices involved in the definition of a background level, we focus instead on the analysis of \hii\ region fluxes without background subtraction throughout this work, but we comment in the text and tables on the effect of the background subtraction, where relevant.

When considering the properties of \hii\ regions we further applied cuts in the BPT (Baldwin-Phillips-Terlevich, \citealt{Baldwin1981, Phillips1986}) diagrams \nii/\ha\ versus \oiii/\hb\ and \sii/\ha\ versus \oiii/\hb, corresponding to the dividing lines of \cite{Kauffmann2003a}, and \cite{Kewley2001}, respectively, to minimise contamination from supernova remnants and planetary nebulae \citep{Groves2023}. This excluded 25\% of the starting sample of nebulae. Finally, we discarded regions overlapping with foreground stars (0.3 \%), or that have a signal-to-noise ratio lower than three in H$\alpha$ or H$\beta$ (0.4\%). We obtain a final sample of 19~901 regions with signal-to-noise ratio larger than three in F2100W. Their distribution across the sample is summarised in Table \ref{tab:tab1}.


\begin{table*}[t]
	\centering
	\begin{tabular}{l c c  c}
	Quantity	   &  Definition  & Derived from dataset & Reference\\[0.15cm]
 \hline 
        $\Sigma_{\rm H_2}$ & $\alpha_{\rm CO}  I_{\rm CO(2-1)}/R_{21},^{\rm a}$  & ALMA & \cite{Leroy2021a} \\[0.15cm]
        E($B-V$) & Eq. \ref{ebv} & MUSE & \cite{Groves2023}  \\[0.15cm]
        $\Sigma_{\star}$ & spectral fitting & MUSE & \cite{Emsellem2022}  \\[0.15cm]
        sSFR & $C_{\rm H\alpha}~I_{\rm H\alpha, corr}/\Sigma_{\star}$,$^{\rm b}$ & MUSE & \cite{Belfiore2023}  \\[0.15cm]
        EW(\ha) & moments of spectra$^{\rm c}$  &  MUSE & \cite{Groves2023}  \\[0.15cm]
        12+log(O/H) & Scal calibration$\rm ^{d}$   &  MUSE & \cite{Groves2023}  \\[0.15cm]
        log(U) & from $\dfrac{[\mathrm{S}\textsc{iii}]\lambda\lambda9069,9532}{[\mathrm{S}\textsc{ii}]\lambda\lambda6717,31}$,$^{\rm e}$  &  MUSE & \cite{Groves2023}  \\[0.15cm]

	\end{tabular}
	\caption{Summary of additional data and physical quantities used in this work. \\
    \textbf{Notes}. \textbf{a}: assuming a Milky Way conversion factor $\alpha_{\rm CO} = 4.35~ \rm M_{\odot} ~pc^{-2} (K~km~s^{-1})^{-1}$ \citep{Bolatto2013} and a CO(2-1)-to-CO(1-0) ratio $R_{21} = 0.65$ \citep{DenBrok2021}. \textbf{b}: $\log{C_{\rm H\alpha}/(\mathrm{M_\odot~yr^{-1}/(erg~s^{-1}}))} = -41.26$, from \cite{Calzetti2007}. \textbf{c}: definition in \cite{Westfall2019}. \textbf{d}: Scal calibration from \cite{Pilyugin2016}. \textbf{e}: ionisation parameter calculated using the calibration from \cite{Diaz1991}.}
	\label{tab:tabfig2}
\end{table*}

\subsection{Balmer decrement and extinction correction}
\label{sec:ebv}

The attenuation correction using the Balmer decrement is computed assuming Case B recombination, temperature $T=10^4$ K and density $n_{\rm e} = 10^2 \, \mathrm{cm}^{-3}$, leading to  $I_{\rm H\alpha, corr}/I_{\rm H\beta, corr} = 2.86$ \citep{Osterbrock2006}. Under these assumptions 
\begin{equation}
E(B-V) = \frac{2.5}{k_{\rm H\beta} - k_{\rm H\alpha}} \log_{10} \left[ \frac{  L_{\rm H\alpha}/L_{\rm H\beta}  } {2.86} \right],
\label{ebv}
\end{equation}
where $k_{\rm H\alpha}$ and $k_{\rm H\beta}$ are the value of the assumed reddening curve at the wavelengths of the two emission lines. We assume the attenuation law of \cite{O'Donnell1994}.

In order to minimise the effect of distance (and correlated distance uncertainties) across our sample, we present results for \hii\ regions in terms of luminosity per unit area, and in the following $I_{\rm H\alpha}$, $I_{\rm H\alpha, corr}$ and $I_{\rm band}$ refer to the luminosity per unit area of observed \ha\ emission, Balmer decrement corrected \ha, and one of the JWST bands, all in units of $\rm erg~s^{-1}~kpc^{-2}$. 

The range of distances covered by the sample (from 5.2 to 19.6 Mpc) may affect the amount of background contamination, although within our small sample we cannot identify a distance-related change in the relations we study in the rest of this work.  Values of luminosity per unit area are corrected by a factor of $\cos{i}$, where $i$ is the galaxy inclination, to account for projection effects (\citealt{Lang2020}, Table \ref{tab:tab1}). Since our sample excludes highly inclined galaxies, not performing this correction has minimal impact on the results.

\subsection{Other ancillary data}

Throughout this work we use a number of additional physical quantities to characterise \hii\ regions and their local environments. We summarise the definitions, data sets leveraged, and the publications where each of these physical quantities were first derived in Table \ref{tab:tabfig2}.

In particular, we define a specific star-formation rate (sSFR) at the location of our \hii\ regions from the attenuation-corrected \ha\ assuming full sampling of the IMF, to allow for a direct comparison with the work of \cite{Belfiore2023}. We define
\begin{equation}
\log{\mathrm{ (sSFR/ yr^{-1})} }= C_{\rm H\alpha}~I_{\rm H\alpha, corr}/\Sigma_{\star},
\label{eq:ssfr}
\end{equation}
where $C_{\rm H\alpha}$ is the conversion factor derived for a fully-sampled initial mass function (IMF) by \cite{Calzetti2007}, and $\Sigma_{\star}$ is the stellar mass surface density estimated from the MUSE data at the location of the \hii\ region, corrected downwards by 0.09 dex to bring it into agreement with the stellar masses obtained by \cite{Salim2018} and \cite{Leroy2019} based on SED fitting (see discussion in \citealt{Belfiore2023}). This stellar mass surface density estimate does not refer to the mass of the cluster ionising the \hii\ region, which cannot be resolved in the MUSE data, but is derived in larger spatial bins, of size comparable to the copt PSF FWHM \citep{Emsellem2022}.

In addition, in Sec. \ref{sec:HII_reg_params} we use the molecular gas mass surface density, as traced by CO(2-1) maps obtained as part of the PHANGS--ALMA program \citep{Leroy2021b, Leroy2021a}, at the positions of our optically defined \hii\ regions. The ALMA data has slightly lower angular resolution than copt ($\sim$ 1.3$''$) for each of our galaxy targets, but considering the integration over the \hii\ region masks the small mismatch in resolution does not affect our results.

\begin{figure*}[ht!]
\centering
\includegraphics[width=0.98\textwidth,trim=0 0 0 10, clip]{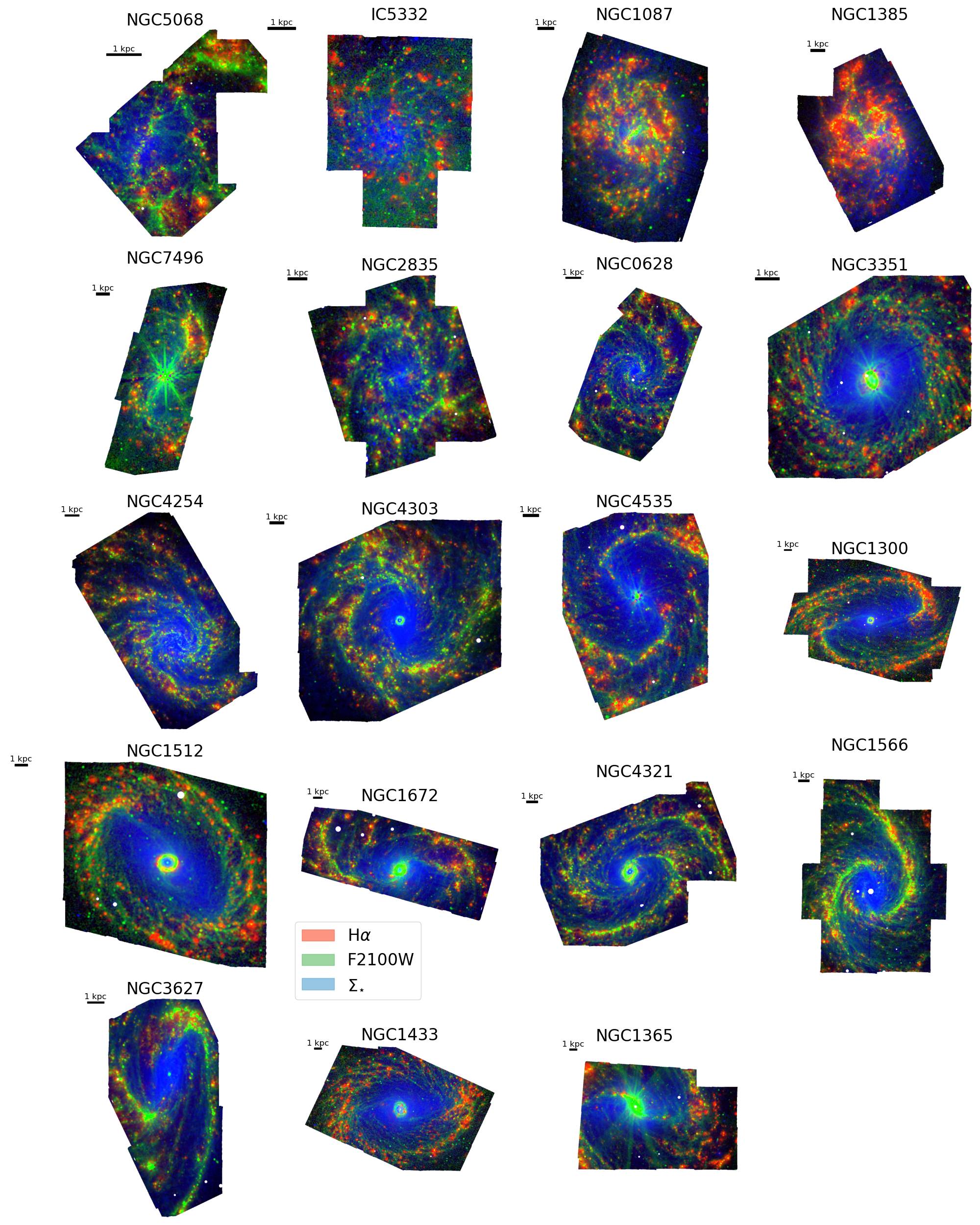}
\caption{Three-colour images for the galaxies in our sample showing a combination of \ha\ emission from MUSE IFS (red), 21~\um\ emission from JWST MIRI F2100W images (green), and stellar mass surface density from MUSE full spectral fitting (blue). The JWST data and MUSE data are shown at matched resolution. Galaxies are shown in order of increasing stellar mass. Foreground stars and the AGN in NGC~1566 and NGC~1365 are masked. The diffraction spikes due to bright AGN in NGC~4535, NGC~7496 and NGC~1365 are shown here but masked in subsequent analysis. The jagged edges of the images are caused by the intersection between the MUSE and MIRI image coverage.}
\label{fig:figAll}
\end{figure*}

\section{Results}
\label{sec:results}

\begin{figure*}[ht!]
\centering
\includegraphics[width=0.7\textwidth,trim=0 30 20 10, clip]{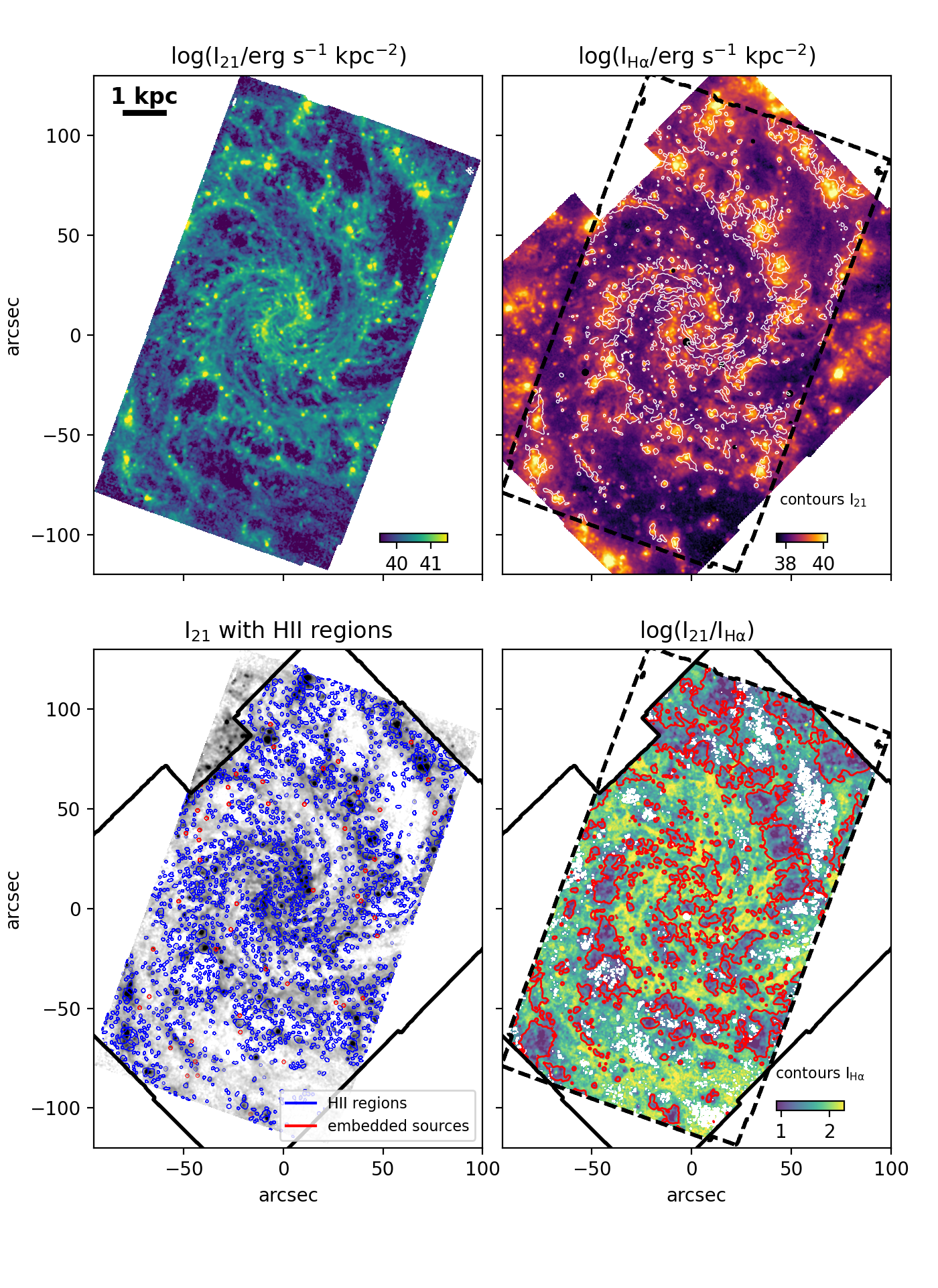}
\caption{Comparison of the \ha\ and 21\um\ emission for NGC~0628. Top-left, 21\um\ surface brightness image obtained using the F2100W filter on MIRI, convolved to the MUSE copt resolution (0.92$''$).
Top-right, map of the \ha\ surface brightness from MUSE \citep{Emsellem2022}.  White contours enclose regions of bright 21~\um\ emission ($I_{21} > 10^{40.5}~ \mathrm{erg ~s^{-1}~ kpc^{-2}}$). Bottom-left: Same as top-left, but with the boundaries of the optically selected \hii\ regions from \cite{Groves2023} in blue and the 21~\um\ sources which do not overlap with optical \hii\ regions from \cite{Hassani2023} in red. Bottom-right: Ratio map of the \ha\ to F2100W surface brightness. Red contours are bright \ha\ regions, $I_{\rm H\alpha} > 10^{38.8} ~ \mathrm{erg ~s^{-1}~ kpc^{-2}}$. White regions within the mapped area represent areas with S/N$<$3 in F2100W. The boundaries of the MUSE (solid black) and MIRI (dashed black) mosaics are shown to aid in visualising the data overlap.}
\label{fig:fig_NGC628}
\end{figure*}

\begin{figure*}[ht!]
\centering
\includegraphics[width=1\textwidth]{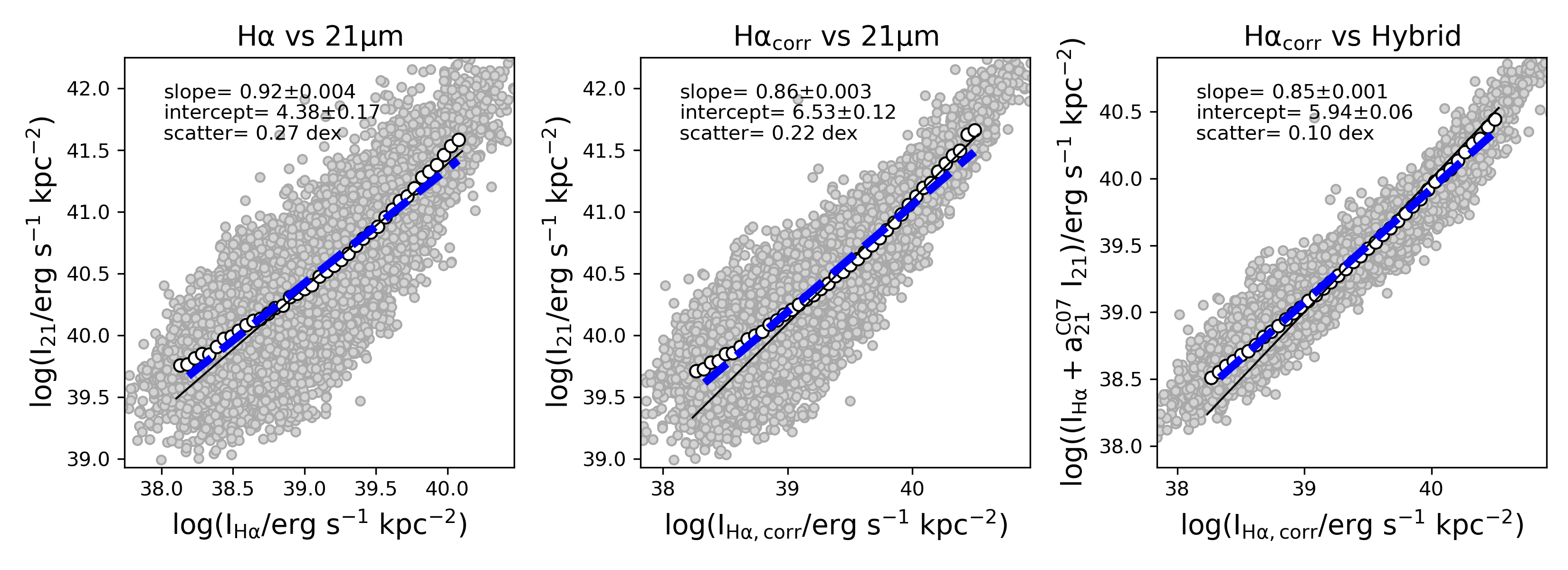}
\caption{Left: $I_{21}$ as a function of $I_{\rm H\alpha}$ for the \hii\ regions (grey points) in our sample. The white dots represent the median relation, while the blue dashed line is the best-fit power law obtained by fitting the data considering the errors associated with the quantities on both axes. The solid black line shows a linear relation normalised to match the data at high surface brightness. The slope and intercept of the best-fit power law together with the scatter with respect to the best model are reported in the top left corner. The meaning of the symbols and lines is the same across the three panels, except that the solid black line the right panel represents a one-to-one relation. Middle: $I_{21}$ as a function of $I_{\rm H\alpha, corr}$. Right: $I_{\rm H\alpha} + a_{21}^{\rm C07} I_{21}$ as a function of $I_{\rm H\alpha, corr}$.
}
\label{fig:fig_HA_21.png}
\end{figure*}

\subsection{Using 21~\um\ emission as a tracer of obscured star formation}
\label{sec:framework}

We aim to provide attenuation-corrected measures of the ionising photon luminosity from massive stars to trace recent star formation in individual \hii\ regions. To do so, we follow the well-established approach of using the MIR to correct for the obscured part of the \ha\ luminosity \citep{Hirashita2003, Calzetti2007, Kennicutt2009, Belfiore2023}. Taking the MIRI F2100W band centred at 21 \um\ as our fiducial band, we can write the attenuation-corrected \ha\ luminosity per unit area as

\begin{equation}
 I_\mathrm{H\alpha, corr}^{\rm hybrid} =   I_{\mathrm{H\alpha}} + a_{\mathrm{21}}~ I_{\mathrm{21}},
\label{eq:hybrid2}
\end{equation}
where all luminosities are expressed in units of $\rm erg~s^{-1}~kpc^{-2}$ (i.e., $I_{\mathrm{21}} = \nu_{21} I_{\nu}$) and $a_{\mathrm{21}}$ is a dimensionless coefficient. This coefficient is generally calibrated empirically because of its complex dependence on star-formation history, star-dust geometry, shape of the dust attenuation law, and potential non-linearity in the dust emissivity with radiation field strength. The dust attenuation at the wavelength of \ha\ can therefore be related to the $I_{21}/I_{\rm H\alpha}$ ratio via
\begin{equation}
A_{\rm H\alpha} = 2.5 \log{ ( 1 + a_{\mathrm{21}}~ I_{21}/I_{\rm H\alpha}) }.
\label{eq:Aha}
\end{equation}
An equivalent formalism can be used with the PAH-tracing bands (Sec. \ref{sec:pah}).

As a starting point for our analysis we consider the numerical value of the coefficient obtained by \cite{Calzetti2007} for a sample of star-forming complexes in nearby galaxies using \textit{Spitzer} 24~\um\ MIPS data: $a_{\rm 24}^{\rm C07}=0.031$. To apply this to JWST data we take into account the different filter bandpasses between MIPS 24~\um\ and MIRI F2100W using the average flux ratio observed across our sample and tabulated by \cite{Leroy2023}, $\langle I_{21}/I_{24} \rangle = 0.91$, leading to $a_{\rm 21}^{\rm C07}=0.034$.



\subsubsection{21 $\rm \mu $m emission in  H\textsc{ii} regions and the diffuse component}

In this section we present the relative spatial distribution of 21~\um\ and \ha\ emission. In Fig. \ref{fig:figAll} we show three-colour images combining \ha\ from MUSE (red) with 21~\um\ emission from MIRI at copt resolution (green) and stellar mass surface density from MUSE full-spectral fitting (blue). Galaxies are ordered by increasing total stellar mass. The figure highlights the variety of stellar and ISM morphologies present in our sample. Throughout the sample we observe a strong spatial correlation between bright 21~\um\ sources and bright \ha\ emission, as expected if both components are powered by young stars (and dust attenuation is moderate, see discussion in \citealt{Leroy2023}).  The fainter 21~\um\ emission outside \hii\ regions appears filamentary, suggesting that such emission traces primarily changes in the column density of cold gas illuminated by the diffuse interstellar radiation field \citep{Leroy2023, Sandstrom2023a}. The presence of diffuse \ha, on the other hand, correlates spatially with the positions of \hii\ region, and can largely be attributed to leaking ionising radiation \citep{Belfiore2022}.

To discuss the relation between \ha\ and 21~\um\ emission more quantitatively we focus on one of the galaxies in our sample, NGC~628, a grand-design spiral galaxy with $\rm log(M_\star/M_\odot) = 10.3$. Fig.~\ref{fig:fig_NGC628} (top row) presents a comparison between the 21~\um\ and \ha\ maps for NGC~628 at matched (copt) resolution.  To quantify the spatial differences between \ha\ and 21~\um\ emission we compute the 21~\um\ flux within and outside the optically defined \hii\ region masks (blue contours in Fig.~\ref{fig:fig_NGC628}, bottom left). We find that in NGC~628, 54\% of the 21~\um\ emission, but only 36\% of the \ha\ emission, lies outside the \hii\ region masks. Across the full sample the fraction of 21~\um\ flux outside \hii\ regions is on average $\sim$ 60\%. This fraction is higher than the fraction of diffuse \ha, which averages to 38\%. 

The large amount of 21~\um\ emission outside \hii\ regions is not due to a population of fully embedded star-forming regions, which may have eluded our optical catalogs. To demonstrate this, we match the MUSE \hii\ region catalog with the catalog of 21~\um\ compact point sources from \cite{Hassani2023} for the four galaxies in common (IC~5332, NGC~628, NGC~1365, NGC~7496). For this sub-sample we identify 21~\um\ sources which do not overlap MUSE \hii\ regions. We allow a maximum overlap of 10\% in area and require sources to have MIR colours typical of the ISM, in order to exclude background galaxies and dusty stars (the selection criteria are discussed in \citealt{Hassani2023}). While the vast majority of 21~\um\ point sources overlap with \hii\ regions, in NGC~628 we find 53 non-overlapping sources, accounting for 0.7\% of the total 21~\um\ emission (shown as red circles in Fig.~\ref{fig:fig_NGC628}, bottom left). These sources do not overlap with supernova remnants or planetary nebulae in the \cite{Groves2023} catalog, so we consider them likely fully embedded star-forming regions. Over the sub-sample of four galaxies in common with \cite{Hassani2023} we find that the fraction of 21~\um\ emission in such embedded point sources is of the order of a few percent. Their contribution is therefore negligible in explaining the diffuse 21~\um\ emission.

The difference in diffuse fraction between \ha\ and 21~\um\ emission implies that areas that are faint in \ha\ show a higher $I_{21}/I_{\rm H\alpha}$ ratio. In the bottom-right panel of Fig.~\ref{fig:fig_NGC628} we show a ratio map of $I_{21}/I_{\rm H\alpha}$ with red contours corresponding to regions of bright \ha\ emission\footnote{$\log{(I_{\rm H\alpha}/\mathrm{erg ~s^{-1} ~kpc^{-2}}} )> 38.8 $, chosen to correspond to the 16$\rm ^{th}$ percentile of the surface brightness distribution of \hii\ regions in the PHANGS-MUSE sample from \cite{Belfiore2022}.}.  The higher $I_{21}/I_{\rm H\alpha}$ in the DIG with respect to \hii\ regions does not imply, however, that the diffuse component suffers from a higher dust attenuation, as may be expected from a naive application of Eq.~\ref{eq:Aha}. In fact, the 21~\um\ and \ha\ emission are powered by different components of the radiation field, and the observed change in the ratio likely reflects differences in the radiation field rather than dust attenuation. The diffuse \ha\ is mostly the result of the leakage of ionising photons from \hii\ regions, with a small (few percent) contribution from old stars \citep{Belfiore2022}. The 21 \um\ emission, on the other hand, is powered by UV and optical light escaping star-forming regions but also originating from the old stellar population, with this ratio changing as a function of local environment \citep{DeLooze2014, Williams2019}. In this work we therefore focus enitirely on the emission properties of \hii\ regions, where both dust and \ha\ should be powered by the same sources. The contamination from the diffuse component is, however, likely to affect the measured properties of the fainter \hii\ regions. A more detailed study of the diffuse infrared emission and its dependence on local environment will instead be presented in a future publication. For the remainder of this work we focus on calibrating the dust correction measures for \hii\ regions.

\begin{figure*}[ht!]
\centering
\includegraphics[width=0.98\textwidth]{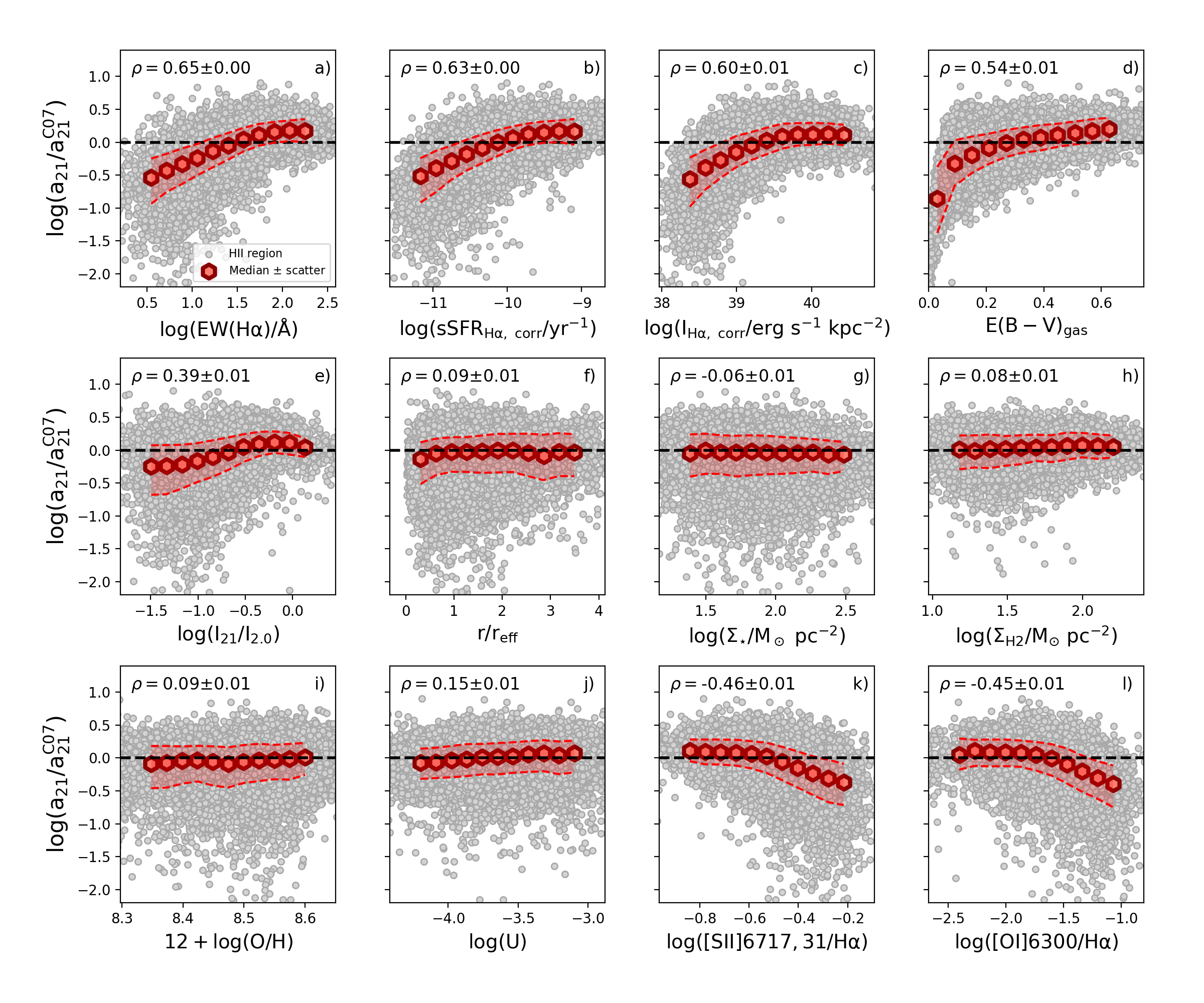}
\caption{The measured coefficient $a_{21}$ normalised to the (rescaled) value from \cite{Calzetti2007} for \hii\ regions as a function of several physical properties. Grey points are individual \hii\ regions, while red hexagons represent the median values as a function of the quantity on the x-axis. In each panel we report the values of the Spearman rank correlation coefficient (and its error). The 16$\rm ^{th}$--84$\rm ^{th}$ percentiles are shown as a red shaded area. The quantities considered on the x-axis are a) equivalent width of \ha\ in emission, b) sSFR surface density, defined as the ratio between the SFR surface density obtained from dust attenuation corrected \ha\ (assuming a fully sampled IMF) and the stellar mass surface density measured from MUSE data via spectral fitting ($\rm \Sigma_{\star}$), c) \ha\ luminosity per unit area corrected for dust attenuation with the Balmer decrement ($I_{\rm H\alpha, corr}$), d) $\rm E(B-V)$ of the gas obtained from the Balmer decrement,
e) ratio of the luminosity per unit area in F2100W to that in F200M ($I_{21}/I_{2.0}$),  f) luminosity per unit area in F2100W, g) MUSE stellar mass surface density from fits to the stellar population to optical spectra, h) molecular gas surface density, derived from ALMA CO(2-1), i) gas-phase metallicity, j) ionisation parameter, k) \sii$\lambda\lambda$6717,31/\ha\ ratio, and l) \oi $\lambda$6300/\ha\ ratio.}
\label{fig:fig_SFR_compare}
\end{figure*}

\subsubsection{H\textsc{ii} region scaling relations at 21 \um}
\label{sec:HII_reg_21}

We test the framework for dust attenuation correction presented in Sect.~\ref{sec:framework} on the scale of \hii\ regions using the JWST F2100W band. We start by presenting the relation between the observed $I_{\rm H\alpha}$ and $I_{\rm 21}$ in \hii\ regions (Fig.~\ref{fig:fig_HA_21.png}). The two quantities show an excellent correlation. A linear fit to log quantities is performed by using the Bayesian framework described in \cite{Kelly2007} which takes into account the errors on both variables. This leads to a sub-linear slope (0.92) and residual scatter around the best-fit relation of 0.27 dex. 

If one corrects the \ha\ emission for dust attenuation via the Balmer decrement, we obtain a relation with lower scatter (0.22 dex), but still a sub-linear slope of 0.86 (Fig.~\ref{fig:fig_HA_21.png}, middle panel). In this space, however, the best-fit relation (blue line) deviates from the observed data for regions of high surface brightness. This result may be compared with \cite{Leroy2023}, who analyse most of the area in each galaxy, therefore including diffuse emission, and obtain a slope of 0.77 (inverse of their reported slope of 1.29 fitting the $I_{\rm H\alpha}$ as a function of $I_{\rm 21}$). 
For \hii\ regions, at $\log(I_{\rm H\alpha} / \mathrm{erg~s^{-1}~kpc^{-2})}> 39.0$ the relation approaches a linear slope (black line). Since bright \hii\ regions also show higher values of dust attenuation, we interpret this as indicating that, when obscured SFR is dominant, both $I_{\rm H\alpha, corr}$ and $I_{\rm 21}$ are good proxies for the total ionising photon production rate. 

Finally, we show the relation between $I_{\rm H\alpha, corr}$ and the hybrid 21~\um\ and \ha\ tracer (Eq. \ref{eq:hybrid2}) using the constant $a_{\rm 21}^{C07}$ value. In this case the quantities on the two axes show correlated errors and we take the non-diagonal elements of the covariance matrix into account to perform the fit, as described in \cite{Kelly2007}. The effect of covariance between the x- and y-axes is small in our case, with an average correlation coefficient of 0.02. Performing a fit to the log quantities, we obtain a power law slope of 0.85 and scatter of the residuals of 0.10 dex. The hybrid tracer therefore provides an excellent match to the Balmer decrement-corrected \ha, but the resulting relation is still not linear. In particular, comparing the best-fit line (dashed blue) with the one-to-one line (black) the calibration leads to an overestimate of the dust correction at low \ha\ surface brightness and a slight underestimate for the high end.

A local background subtraction leads to relations that are closer to linear, but suffer from larger scatter (e.g. a slope of 1.03 for the relation between $I_{\rm H\alpha, corr}$ and the hybrid 21 \um\ and \ha, and a scatter of 0.37 dex). The deviation from linearity at low \ha\ surface brightness is less evident, but still present. Because of the difficulty to provide a robust measure of the background, especially for faint regions, we do not further pursue this calculation.

\subsubsection{Secondary dependencies of the calibration coefficient for 21 \um\ emission }
\label{sec:HII_reg_params}

In this section we test whether an additional physical parameter drives the deviation from linearity observed using a constant $a_{21}$ coefficient. We assume that the Balmer decrement traces dust attenuation across our sample (i.e. that we do not see any fully embedded emission) and calculate the $a_{21}$ coefficient using equation \ref{eq:Aha}. For ease of benchmarking we consider the ratio between the inferred $a_{21}$ and the C07-equivalent value of $a_{21}^{\rm C07}$.

Fig.~\ref{fig:fig_SFR_compare} shows $a_{21}/a_{21}^{\rm C07}$ as a function of several \hii\ region properties derived either in this work or in \cite{Groves2023}.
In several of the panels in Fig.~\ref{fig:fig_SFR_compare} the quantities on the x- and y-axis are not statistically independent. We have checked, however, that analytical propagation of the error correlations implies a degree of correlation much smaller (on average 2 dex smaller) than that observed in the data: i.e. the observed trends cannot be explained by covariance among the errors alone. 

We quantify the correlations between each set of quantities via the Spearman rank correlation coefficient. In order to estimate the error on this statistic we perform a Monte Carlo simulation by adding Gaussian noise to the data. We take into account the correlations between the errors on the x- and y-axis by sampling from a 2D Gaussian with the appropriate covariance matrix for each data point. 
The median and standard deviations of 1000 Monte Carlo runs are taken as our best estimates of the Spearman rank correlation coefficient and its error, and are shown in the top-left for each panel in Fig.~\ref{fig:fig_SFR_compare}. Given the relatively high signal-to-noise ratio in \hii\ regions and the large dynamic range spanned in most quantities with respect to their errors, the errors on the correlation coefficients are found to be small ($\sim 0.01$ in all panels).

The inferred $a_{21}$ coefficient demonstrates significant positive correlations with EW(\ha) ($\rho = 0.65$, panel a), sSFR estimated from the attenuation-corrected \ha\ ($\rho = 0.63$, panel b), attenuation-corrected \ha\ surface brightness ($I_{\rm H\alpha,  corr}$, $\rho = 0.60$, panel c),  E(B$-$V) computed from the Balmer decrement ($\rho = 0.54$, panel d), and the ratio of 21 \um\ to 2.0 \um\ surface brightness, where the surface brightness at 2.0 \um\ is taken from the NIRCam imaging data in the F200W filter ($\rho = 0.39$, panel e). EW(\ha), sSFR, and $I_{21}/I_{2.0}$ trace the luminosity of the \hii\ region with respect to that of the old stars. E(B$-$V) is known to correlate with both attenuation-corrected \ha\ (e.g. \citealt{Emsellem2022}), and EW(\ha) \citep{Groves2023}. Overall, these trends suggest that $a_{21}$ is slightly higher than the \cite{Calzetti2007} value for the brightest (and dustiest \hii\ regions), while it should be significantly lower, by up to 1.0~dex for faint, less dusty regions. All these correlations flatten for the brightest \hii\ regions.

Substantially weaker correlations (with $|\rho| < 0.3$) are found as a function of galactrocentric radius (panel f), stellar mass surface density (panel g), molecular gas surface density (panel h), gas-phase metallicity (panel i), or ionisation parameter (the ratio between the ionising photon flux and the hydrogen density in \hii\ regions, panel j). 

Finally, the relation between $a_{21}$ and \sii $\lambda\lambda$6717,31/\ha\ (panel k) or \oi $\lambda$6300/\ha\ (panel l) is flat at low values of these line ratios, but shows a strong negative correlation ($\rho = -0.46$ and $\rho = -0.45$, respectively) for higher ones.  
These line ratios have been studied in our sample by \cite{Belfiore2022} and are found to increase with decreasing \ha\ surface brightness, corresponding to the transition between \hii\ regions and diffuse ionised gas \citep{Haffner2009, Zhang2017}. Even though we are focusing on \hii\ regions here, we are likely seeing the increased contribution of the diffuse ionised gas background to the flux within the masks corresponding to faint \hii\ regions going hand in hand with an increase in the relative significance of the 21~\um\ background. 

We have tested the effect of defining a local background subtraction, which leads to an increase in the scatter and a decrease in the strength of all correlation coefficients, but does not change the shape and sign of the strongest ones. 

\subsection{Using PAH-tracing bands to measure obscured star formation}
\label{sec:pah}

In this section we test the ability of PAH-dominated bands to trace the obscured component of star formation. We consider the continuum-subtracted F335M NIRCam band (F335M$\rm _{PAH}$), and the MIRI bands centred at 7.7~\um\ (F770W), 10~\um\ (F1000W), and 11.3~\um\ (F1130W). Emission in the 7.7~\um\ and 11.3~\um\ bands is dominated by features generally associated with PAHs \citep{Smith2007, Draine2007}, while the star-light contamination in the F335M band is removed as described in Sec. \ref{jwst_data} and \cite{Sandstrom2023a}. The nature of the continuum emission underlying the PAH features and dominating the 10 \um\ band emission remains unclear \citep{Smith2007, Li2020}. \cite{Leroy2023} find that the 10 \um\ continuum  closely follows the PAH-dominated bands, rather than the 21 \um\ emission. In this work we therefore refer collectively to F335M$\rm _{PAH}$, F770W, F1000W, and F1130W as `PAH-tracing' bands.

\begin{table}[]
	\centering
	\begin{tabular}{l c c c c}
	\\
	\multicolumn{4}{c}{$\log{(I_{\rm band}/I_{21})}$} \\
    Band &   Median      &   16$\rm ^{th}$ perc. & 84$\rm ^{th}$ perc.  \\
	\hline
    F335M$\rm _{PAH}$ & $-$0.41 & $-$0.62 & $-$0.24 \\  
	F770W & 0.65 & 0.54 & 0.73 \\  
	F1000W & 0.10 & $-$0.04 & 0.18 \\  
	F1130W & 0.60 & 0.45 & 0.68 \\  
	\end{tabular}
	\caption{Median (and 16$\rm ^{th}$ and 84$\rm ^{th}$ percentiles) of the ratio between the F335M$\rm _{PAH}$, F770W, F1000W, F1130W to the F2100W luminosity (in units of $\rm erg~s^{-1}~kpc^{-2}$) for the \hii\ regions in our sample. The dependence of the $I_{\rm band}/I_{21}$ on EW(\ha) is shown in Fig. \ref{fig:fig_PAH_21.png}.}
	\label{tab:PAH_ratios}
\end{table}

\begin{figure*}[]
\centering
\includegraphics[width=\textwidth]{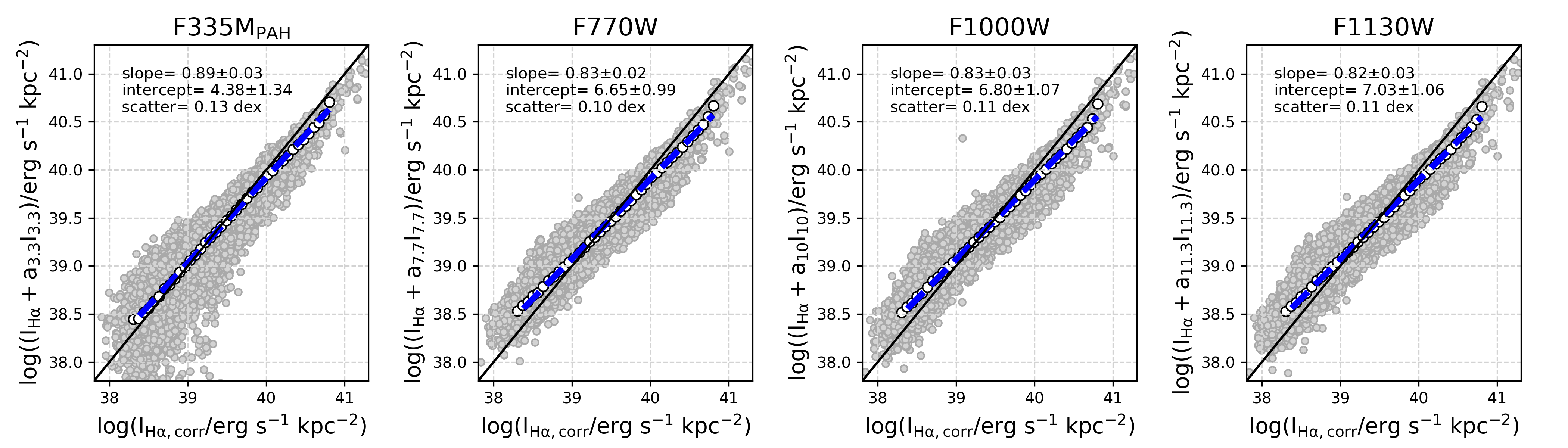}
\caption{Comparison of \ha\ hybridised with F335M$\rm _{PAH}$, F770W, F1000W and F1130W and the attenuation-corrected \ha\ obtained using the Balmer decrement. The dashed blue line is the best linear fit while the white circles represent a binned average. The slope and intercept of the best-fit power law together with the scatter with respect to the best model are reported on the top left corner. The solid black line represents the one-to-one relation.}
\label{fig:calib_PAH}
\end{figure*}

\subsubsection{PAH-tracing band ratios in H\textsc{ii} regions}

We adopt an empirical strategy to set fiducial values of the hybridisation coefficients for each band, $a_{\rm band}$. We calculate these coefficients, which we refer to as `C07-equivalent', by scaling the value of the \cite{Calzetti2007} 21~\um\ coefficient using the average band ratio between PAH-tracing bands and F2100W. In detail, we define
\begin{equation}
    \log{(a_{\rm band}^{\rm C07})} \equiv  \log{(a_{21}^{\rm C07})} - \langle \log{(I_{\rm band}/I_{21})} \rangle,
    \label{eq:PAH_corr}
\end{equation}
where $\langle\log{(I_{\rm band}/I_{21})}\rangle$ is the median band ratio between the bands of interest (3.3, 7.7, 10, and 11.3 \um) and 21 \um\ in our \hii\ region sample. We report the median, 16$\rm ^{th}$, and 84$\rm ^{th}$ percentiles of the distribution of $I_{\rm band}/I_{21}$ in Table \ref{tab:PAH_ratios}. The scatter in the $I_{\rm band}/I_{21}$ lies in the range of 0.2~dex (F770W) to 0.4~dex (F335M$\rm _{PAH}$). Part of this larger scatter for F335M$\rm _{PAH}$ may be attributed to residuals in the continuum subtraction.


\subsubsection{H\textsc{ii} region scaling relations for PAH-tracing bands}

Fig. \ref{fig:calib_PAH} shows the relations between \ha\ corrected using the Balmer decrement and the \ha\ hybridised with each of the PAH-dominated bands using the $a_{\rm band}^{\rm C07}$ factor defined in Eq.~\ref{eq:PAH_corr}. For all bands considered we obtain a best-fit power law with sub-linear slopes (ranging from 0.82 to 0.89) and $\sim$0.10~dex scatter (except for F335M$\rm _{PAH}$ for which we obtain a marginally larger scatter of 0.13 dex). 

These results are remarkably consistent with those obtained with F2100W in Sec. \ref{sec:HII_reg_21}. In fact, Fig. \ref{fig:calib_PAH} demonstrates that the hybrid recipe overestimates the dust correction at low surface brightness levels ($\rm log(I_{H\alpha, corr}/erg~ s^{-1}~kpc^{-2}) < 39$), while underestimating it at high surface brightness levels, as already observed for F2100W.

\subsubsection{Secondary relations of $a_{\rm band}$ for PAH-tracing bands}
\label{Sec:seconadary_PAH}

\begin{figure*}[ht!]
\centering
\includegraphics[width=0.8\textwidth]{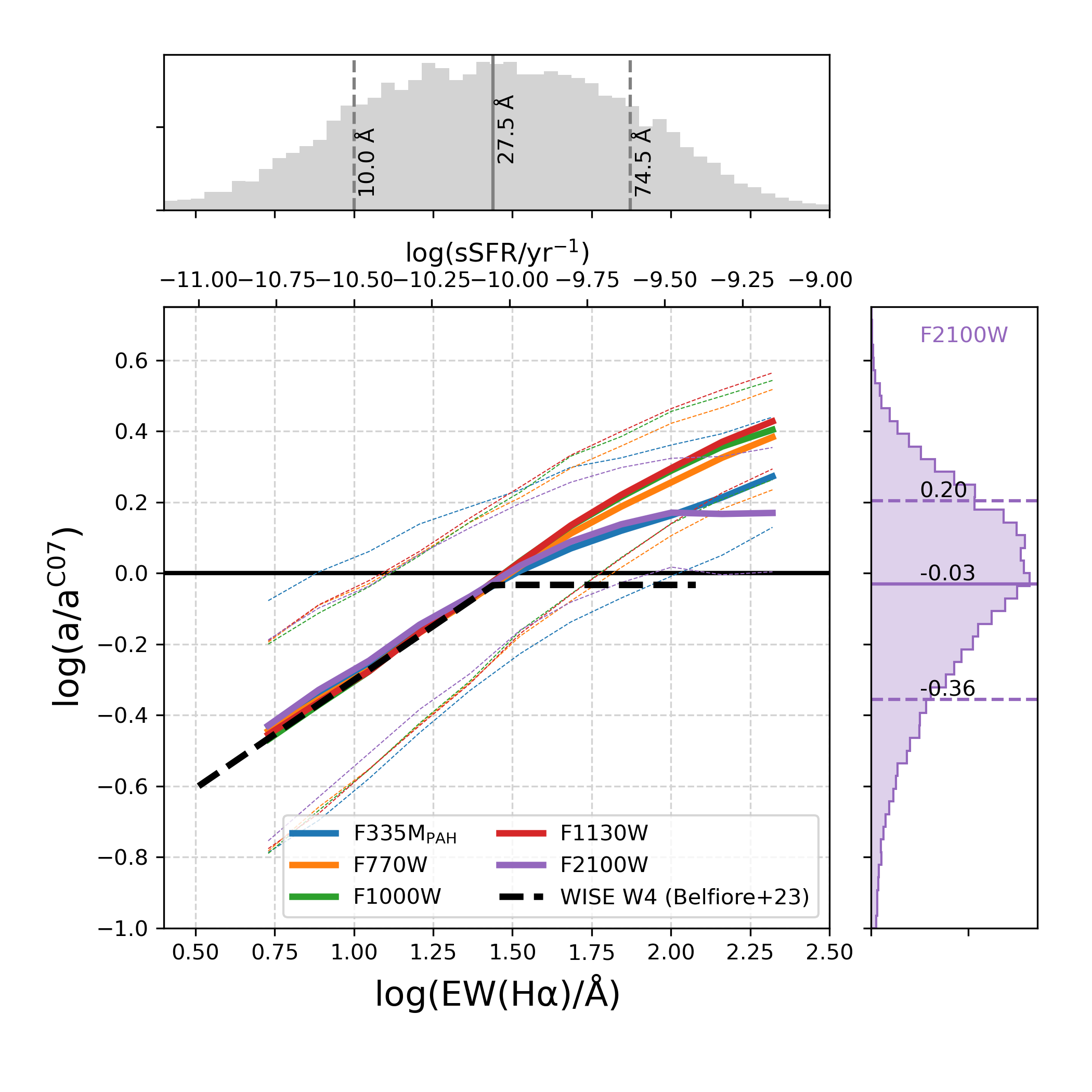}
\caption{$\log{a_{\rm band}}$ normalised to the C07-equivalent values as a function of EW(\ha) and sSFR (alternative x-axis on top) for \hii\ regions in our sample, where band = [F335M$_{\rm PAH}$ (blue), F770W (orange), F1000W (green), F1130W (red), F2100W (purple)]. The coloured solid lines represent the median relations, while the coloured dashed lines show the 16$\rm ^{th}$ and 84$\rm ^{th}$ percentiles of the distribution. The black dashed line is the best-fit to the low-resolution (kpc-scale) \ha\ + WISE W4 data from \cite{Belfiore2023}. The grey histogram (top) shows the distribution of EW(\ha) in the \hii\ regions used in this work (with 16$\rm ^{th}$, 50$\rm ^{th}$, 84$\rm ^{th}$ percentiles marked and labelled). The purple histogram (right) shows the distribution of $\log{ a_{\rm 21}/a_{\rm 21}^{\rm C07} }$ for the region in our sample, with the 16$\rm ^{th}$, 50$\rm ^{th}$, 84$\rm ^{th}$ percentiles marked and labelled. }
\label{fig:fig6}
\end{figure*}

\begin{table*}[]
	\centering
	\begin{tabular}{l c c c c c c}
	\\
       &         &  F335M$_{\rm PAH}$  &  F770W & F1000W & F1130W & F2100W \\
	\hline \\[-0.3cm]
    constant coeff. C07 & $a_{\rm band}^{\rm C07}$ &   0.078 &  0.0077 & 0.027 & 0.0087 & 0.034 \\[0.15cm]
    median measured & $a_{\rm band}$ &   0.087$^{+0.07}_{-0.05}$ &  0.0072$^{+0.007}_{-0.004}$ & 0.026$^{+0.027}_{-0.015}$ & 0.0084$^{+0.009}_{-0.005}$ & 0.031$^{+0.023}_{-0.017}$ \\[0.15cm]
    median measured (backg. subtracted) & $a_{\rm band}^{\rm bkg~sub}$ &   0.13$^{+0.16}_{-0.08}$ &  0.016$^{+0.03}_{-0.01}$ & 0.06$^{+0.1}_{-0.04}$ & 0.020$^{+0.03}_{-0.014}$ & 0.051$^{+0.07}_{-0.03}$ \\[0.15cm]
	\end{tabular}
	\caption{Hybridisation coefficients for various JWST bands. The top row shows the value of the \cite{Calzetti2007} coefficient, originally derived from Spitzer 24\um, re-scaled using the median band ratios (see text). The second and third rows show the median $a_{\rm band}$ obtained in our sample of \hii\ regions before and after applying a local background subtraction.   }
	\label{tab:recommend}
\end{table*}

In Fig. \ref{fig:fig6} we show the median trends of the $a_{\rm band}$ coefficients for different bands as a function of EW(\ha). The values of $a_{\rm band}$ are scaled logarithmically and normalised, so that the zero on the y-axis corresponds to the C07-equivalent value. The trend for F2100W (purple) was already presented in Fig. \ref{fig:fig_SFR_compare}, but here we compare it directly with the behaviour of the other bands. For $\log{\rm EW(\ha)/\AA} <1.5$ (or $\log(\rm sSFR/yr^{-1}) <-10$) all bands follow a closely matching trend of increasing $a_{\rm band}$ with EW(\ha), reaching the C07-equivalent value at $\log{\rm EW(\ha)/\AA} \sim 1.5$. 

At high EW(\ha), on the other hand, the behaviour of different bands diverges. The dependence of $a_{21}$ on EW(\ha) flattens for $\log{\rm EW(\ha)/\AA} > 2$, plateauing to a constant value of $a_{21}$ that is 0.16~dex higher than the C07-equivalent one. The F335M$_{\rm PAH}$ band also shows a flattening in the slope of the relation but with no plateau, while F770W, F1000W, and F1130W follow increasing trends. 

The difference between the PAH-tracing bands reflects the relative brightness of different features in the MIR spectrum. For example, \cite{Chastenet2023a} find that the ratio of $(I_{7.7}+I_{11.3})/I_{21}$, which traces PAH abundance \citep{Draine2021}, is relatively constant in the diffuse ISM, but decreases in bright \hii\ regions. \cite{Egorov2023} find that this ratio  in \hii\ regions anti-correlates with the ionisation parameter, which is tracing the density of the extreme UV photons in the region. Moreover, \cite{Chastenet2023b} find that the F335M$\rm _{PAH}$ band is enhanced and the F1130W band is suppressed in regions of high \ha\ surface brightness. These trends indicate that the abundance of PAHs is lowered in \hii\ regions, due to their destruction by extreme UV photons, but also imply changes in the average properties of the PAH population, namely a decrease in average size and an increased abundance of charged grains and/or hotter grains.


\begin{figure}[ht!]
\centering
\includegraphics[width=0.5\textwidth]{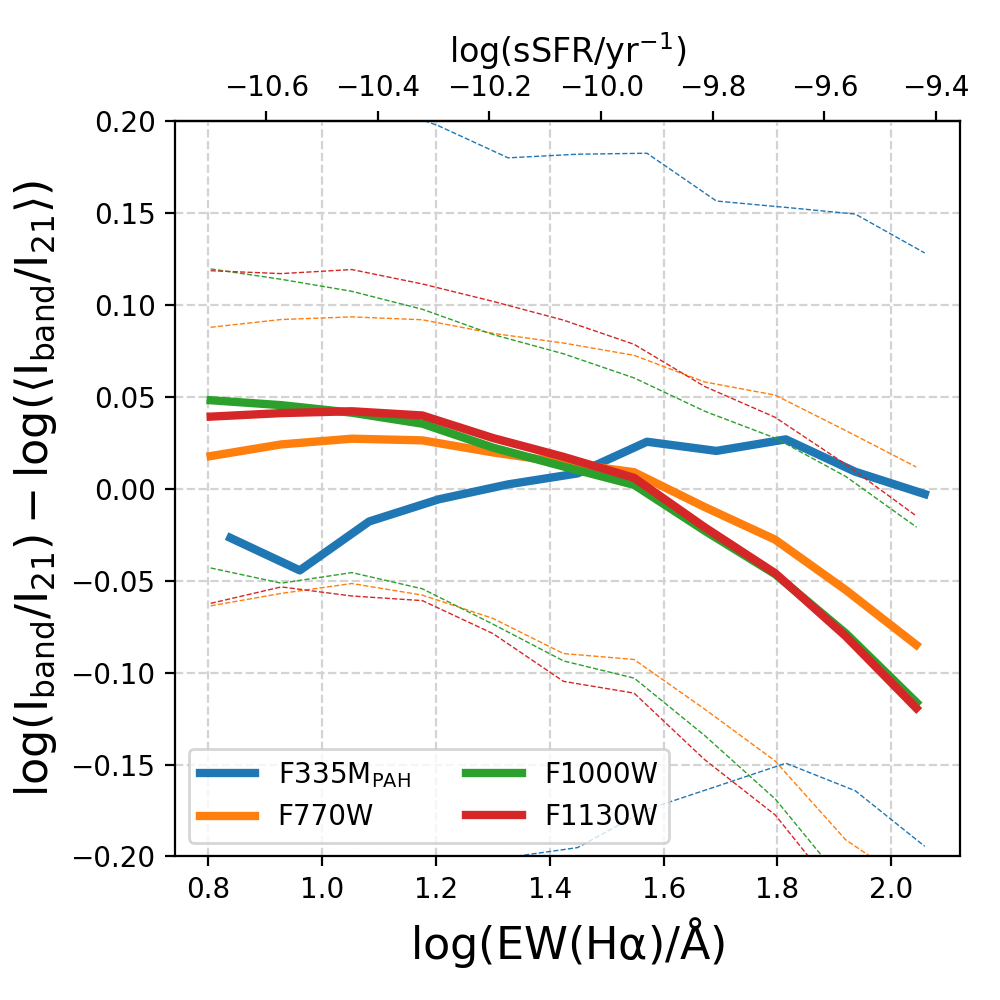}
\caption{The ratio of $\rm I_{band}/I_{21}$ normalised to the sample mean ($\rm \langle I_{band}/I_{21} \rangle$) as a function of EW(\ha), for band= [F335M$_{\rm PAH}$ (blue), F770W (orange), F1000W (green), F1130W (red)]. The solid lines represent the median, while the dashed lines correspond to the 16$^{\rm th}$ and 84$^{\rm th}$ percentiles. The alternative x-axis on top shows the sSFR associated with each EW(\ha) value. The median trends show that $ I_{3.3}/I_{21}$ does not show the same decrease with EW(\ha) as the ratios of the other PAH bands. }
\label{fig:fig_PAH_21.png}
\end{figure}

We show the change in the $I_{\rm band}/I_{21}$ ratios as a function of EW(H$\alpha$) in Fig. \ref{fig:fig_PAH_21.png}. The ratio for each band is normalised by subtracting the median value for the sample, allowing for easier comparison of the trend across bands. Fig. \ref{fig:fig_PAH_21.png} shows the expected decrease in the ratios of $I_{7.7}/I_{21}$, $I_{10}/I_{21}$, and $I_{11.3}/I_{21}$ as a function of EW(H$\alpha$). $I_{7.7}/I_{21}$ shows a slightly shallower slope than $I_{10}/I_{21}$ and $I_{11.3}/I_{21}$. The $I_{3.3}/I_{21}$ ratio, on the other hand, shows a much flatter (and even increasing) trend.
 
Finally, our targets are relatively metal-rich, so we do not probe the low-metallicity regime typical of dwarf galaxies where the PAH fraction is observed to be lower \citep{Madden2006, Draine2007, Khramtsova2013}. Previous work, however, focused on studying entire galaxies or kpc-size regions. On the scale of \hii\ regions in our metallicity range the PAH fraction is found not to depend on metallicity \citep{Egorov2023}, but rather on the intensity of hydrogen-ionizing radiation that probably regulates the PAH destruction in \hii\ regions. Overall EW(\ha) remains the most significant secondary correlation for the PAH bands, and we therefore do not consider additional ones in this work.



\section{Discussion}
\label{sec:discussion}

This work presents an unprecedented set of resolved IR data for star-forming regions at $\sim$ 100~pc scales. Our results show the promise, and caveats, involved in the use of hybrid SFR indicators on the scales of \hii\ regions.
In particular, we use IR observations with JWST at 3.3, 7.7, 10, 11.3, and 21 \um\ to correct \ha\ emission from \hii\ regions for dust attenuation using a hybrid recipe. Using a scaled version of the C07 coefficient (Eq. \ref{eq:PAH_corr}) leads to hybrid SFR estimations which agree with the SFR derived via the Balmer decrement to within $\sim$0.1 dex, but show systematic deviations for both low and high SFR. These deviations correlate most notably with quantities tracing sSFR (e.g., EW(\ha)). 

\subsection{Recommendation for using constant $a_{\rm band}$}

We report in Table \ref{tab:recommend} the median values of the $a_{\rm band}$ coefficients for our full sample of 19~901 \hii\ regions, together with the C07-equivalent values. Our median $a_{\rm band}$ are in excellent agreement with the C07-equivalent values, but the scatter around these values (quantified as the 16$\rm^{th}$ and 84$\rm^{th}$ percentiles) are large.

In Table \ref{tab:recommend} we also present the median values of the $a_{\rm band}$ coefficients after performing a local background subtraction. These values are $\sim$ 0.2 to 0.4~dex larger than our nominal ones. The background subtraction more strongly affects the F770W, F1000W, and F1130W bands ($a$ increases by $\sim$ 0.4 dex), while it has a smaller effect on F2100W and F335M$_{\rm PAH}$ ($a$ increases by $\sim$ 0.2 dex).

The agreement between our values and the C07-equivalent ones is not immediately expected.  \cite{Calzetti2007} estimate the background in large sub-galactic regions, but not using local annuli, therefore likely resulting in an intermediate estimate of the value of the background level. However, the median $a_{\rm band}$ will also depend on the distribution of EW(\ha) in the sample of regions considered and on the median spatial resolution. 

In fact, Fig. \ref{fig:fig6} demonstrates that the plateau observed using kpc-resolution IR data (WISE 22\um\, in \citealt{Belfiore2023}) is a resolution effect, due to the blending of different \hii\ regions with each other and with the diffuse medium. The trend observed by \cite{Belfiore2023} using kpc, 15$''$-resolution WISE W4 22~\um\ data and the full PHANGS--MUSE galaxy sample, is shown as the black dashed line in Fig. \ref{fig:fig6}. 
JWST F2100W observations, providing a factor of $\sim$ 15 higher spatial resolution than WISE W4, reveal that the relation between $a_{21}$  and EW(\ha) plateaus at higher EW(\ha) (and $a_{21}$) than seen in the low-resolution data. Since our observations resolve the inter-cloud distance ($\sim$ 100~pc, \citealt{Kim2022a}), and the contamination from the diffuse background is minimal for the bright regions, we expect the trend uncovered by JWST to persist at even higher spatial resolution. This hypothesis can be tested with upcoming JWST observations of more nearby (or Local Group) galaxies.


Finally, we note that our background subtraction strategy does not remove the trend of increasing $a_{\rm band}$ with EW(\ha). Future work may consider more advanced approaches to model the background and test whether such contamination can explain the behaviour observed in Fig. \ref{fig:fig6} at low EW(\ha).

\begin{figure}[t!]
\centering
\includegraphics[width=0.48\textwidth]{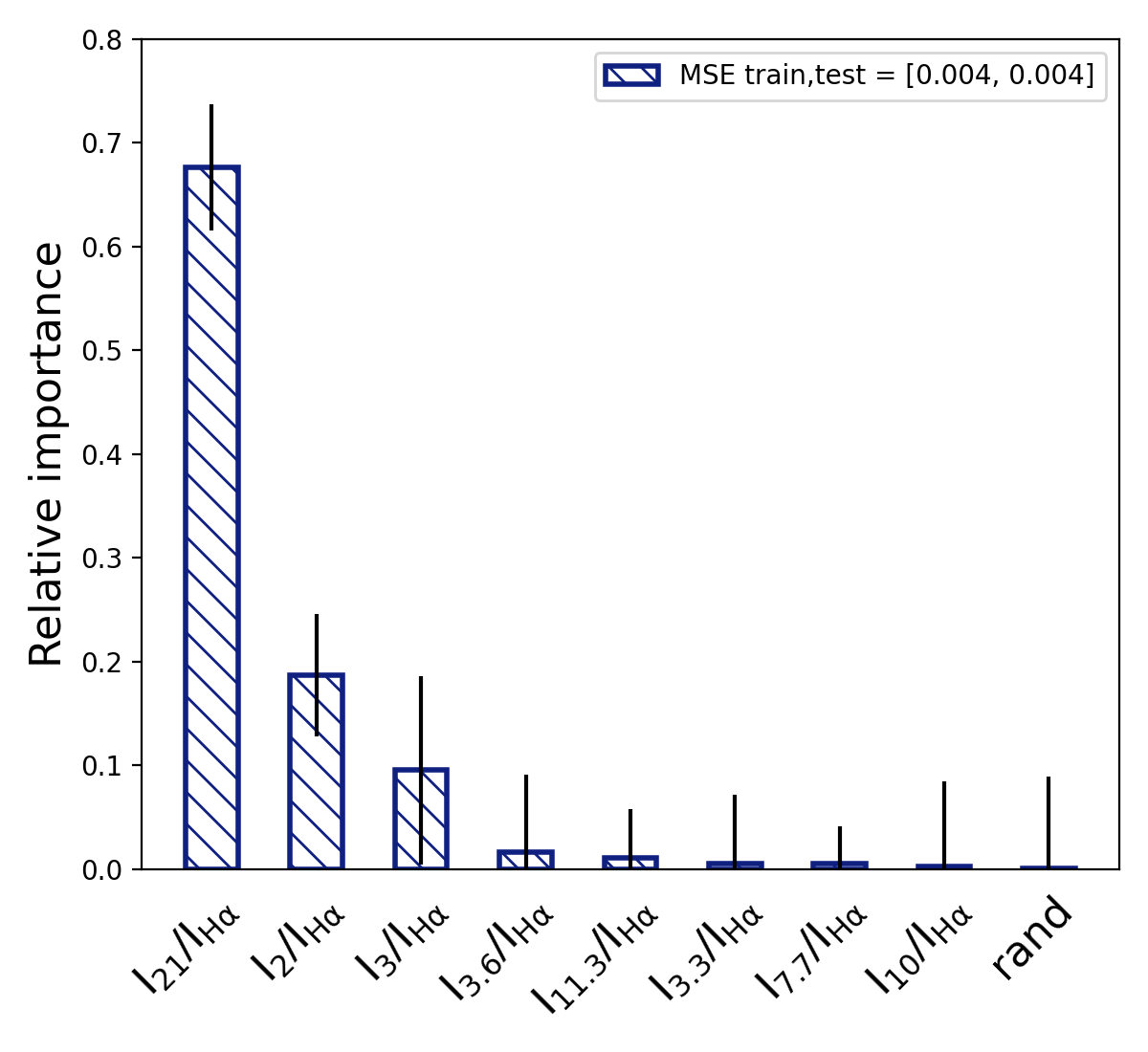}
\caption{Relative importance of different band ratios in predicting the E(B$-$V) of \hii\ regions using a random forest model. The error-bars are obtained by bootstrap random sampling. The mean squared errors (MSE) are shown for both the testing and the training set. $I_{21}/I_{\rm H\alpha}$ and $I_{2.0}/I_{\rm H\alpha}$ account for most of the variance in the data.}
\label{fig:fig7}
\end{figure}

\subsection{Modelling the impact of old stellar population on $\rm a_{band}$}

The trend of increasing $a_{\rm band}$ with EW(\ha) likely reflects the increasing contribution from dust heated by old stellar populations in fainter regions \citep{Cortese2008, Leroy2012, Boquien2016, Nersesian2019}, and, for the PAH-tracing bands, the roughly constant band ratios in diffuse regions \citep{Chastenet2019}. 

Modelling the behaviour of $a_{\rm band}$ as a function of EW(\ha), or one of the other parameters studied in Sec. \ref{sec:HII_reg_params}, would provide a more accurate estimate of the dust-attenuated star formation, and a prescription more readily applicable to lower-resolution data. Several authors have presented calibrations including  such additional terms, for example colours, sSFR, or $\rm M_\star$ \citep{Boquien2016, Zhang2023, Belfiore2023, Kouroumpatzakis2023}.  

We aim to provide such a second-order correction applicable to nearby galaxies data. The functional form that this correction should take is not evident a priori. For the sake of simplicity, here we add an additional linear term to the conventional hybrid star formation recipe presented in Eq. \ref{eq:hybrid2}, leading to 
\begin{equation}
I_{\rm H\alpha, corr} = I_{\rm H\alpha} + \alpha~ I_{\rm IR} + \beta~ I_Q,
\label{eq:plane_fit }
\end{equation}
where $I_{\rm IR}$ is one of the JWST IR dust-tracing bands, and $I_Q$ is a second physical quantity appropriately chosen to best reproduce the extinction correction from the Balmer decrement. As before, we express all surface brightness measurements in units of $\rm erg~s^{-1}~kpc^{-2}$.

We first consider the task of deriving an estimate for the obscured SFR using two JWST bands: a dust-tracing band (F2100W or any of the PAH-tracing bands), and a second band tracing stellar continuum emission (F200W, F300M or F360M) to account for the dust heating from old stars. To determine the best combination of bands for such a calibration we run a random forest regression, using the algorithm implementation in \textsc{scikit-learn} \citep{Pedregosa2011}. In particular, we use the random forest model to determine the relative importance of input features \citep{Bluck2022, Baker2022}. This approach is conceptually similar to more traditional principal component analysis, but allows the generation of more complex, non-linear models.

The random forest algorithm is a form of supervised machine learning that builds a non-parametric model using multiple decision trees, trained to decrease the mean square error at each split. We use E($B-V$) as the target variable since the goal is to understand how best to perform the dust correction, and $I_{\rm band}/I_{\rm H\alpha}$, where `band' is one of the JWST bands available to us (F200W, F300M, F360M, F335M$_{\rm PAH}$, F770W, F1000W, F1130W, F2100W), as input features. We also consider an array of random numbers as an additional input feature to test the performance of the algorithm. The dataset is subdivided into a training (80\% of the full sample) and testing (remaining 20\%) subset of \hii\ regions. Only \hii\ regions covered by both NIRCam and MIRI and with signal-to-noise ratio grater than 3 in all JWST bands are used for this computation.
After training, the mean square error (MSE) of the test set is compared with that of the training sample in order to check the quality of the resulting model and avoid overfitting. 

We find that the combination of $I_{21}/I_{\rm H\alpha}$ and $I_{2.0}/I_{\rm H\alpha}$ accounts for nearly all the variance in E(B$-$V) (Fig. \ref{fig:fig7}). Once these two variables are taken into account, the remaining ones have residual importance consistent with random within the errors. The MSE of the training and test set are comparable (reported on the top right of Fig. \ref{fig:fig7}), demonstrating that the algorithm produced a successful model. This analysis confirms two of the key insights from this work: 1) F2100W is the best band to use for hybridisation among the bands tracing dust emission available in our PHANGS--JWST filter set, and 2) there exists an evident secondary dependence on the stellar mass-tracing NIRCam bands, of which the best is F200W.

Having established $I_{21}$ and $I_{2.0}$ as the two best variables, we fit the extinction-corrected \ha\ surface brightness with a two-parameter regression model, using the Bayesian formalism discussed in \cite{Hogg2010a}. We fit for the $\alpha$ and $\beta$ parameters in Eq. \ref{eq:plane_fit } and for intrinsic scatter in the relation, using the Bayesian Monte Carlo sampler \textsc{emcee}, and obtain best-fit as relation 
\begin{equation}
I_{\rm H\alpha, corr} = I_{\rm H\alpha} + 0.025 ~ I_{21} - 3.0\times10^{-4}~ I_{2},
\label{eq:plane_fit2 }
\end{equation} 
We also fit using the F300M and F360M NIRCam bands, obtaining respectively $\alpha = 0.025$, $\beta=8\times10^{-4}$, and $\alpha =0.025 $, $\beta=7\times10^{-4}$.

Our approach here is similar to the one of \cite{Kouroumpatzakis2023}, who use spectral energy distribution models generated with the code \textsc{cigale} to study how best to determine SFR from JWST bands. In their work, however, \cite{Kouroumpatzakis2023} do not consider $I_{\rm H\alpha}$, but model the SFR as a linear combination of $I_{21}$ and $I_{2.0}$. Their results are therefore not directly comparable. We leave a direct comparison of our results with spectral energy distribution models to future work.

Finally, we repeat our two-parameter fit using stellar mass surface density $\Sigma_{\star}$, which may be used in the absence of JWST NIRCam data. We obtain 
\begin{equation}
I_{\rm H\alpha, corr} = I_{\rm H\alpha} + 0.028~ I_{21} -0.085 \times 10^{31}~ \Sigma_{\star},
\label{eq:calib_star}
\end{equation}
where surface brightnesses are in units of $\rm erg^{-1}~s^{-1}~kpc^{-2}$ and $\Sigma_{\star}$ is in units of $\rm M_\odot~kpc^{-2}$.

\begin{figure*}[t!]
\centering
\includegraphics[width=0.7\textwidth]{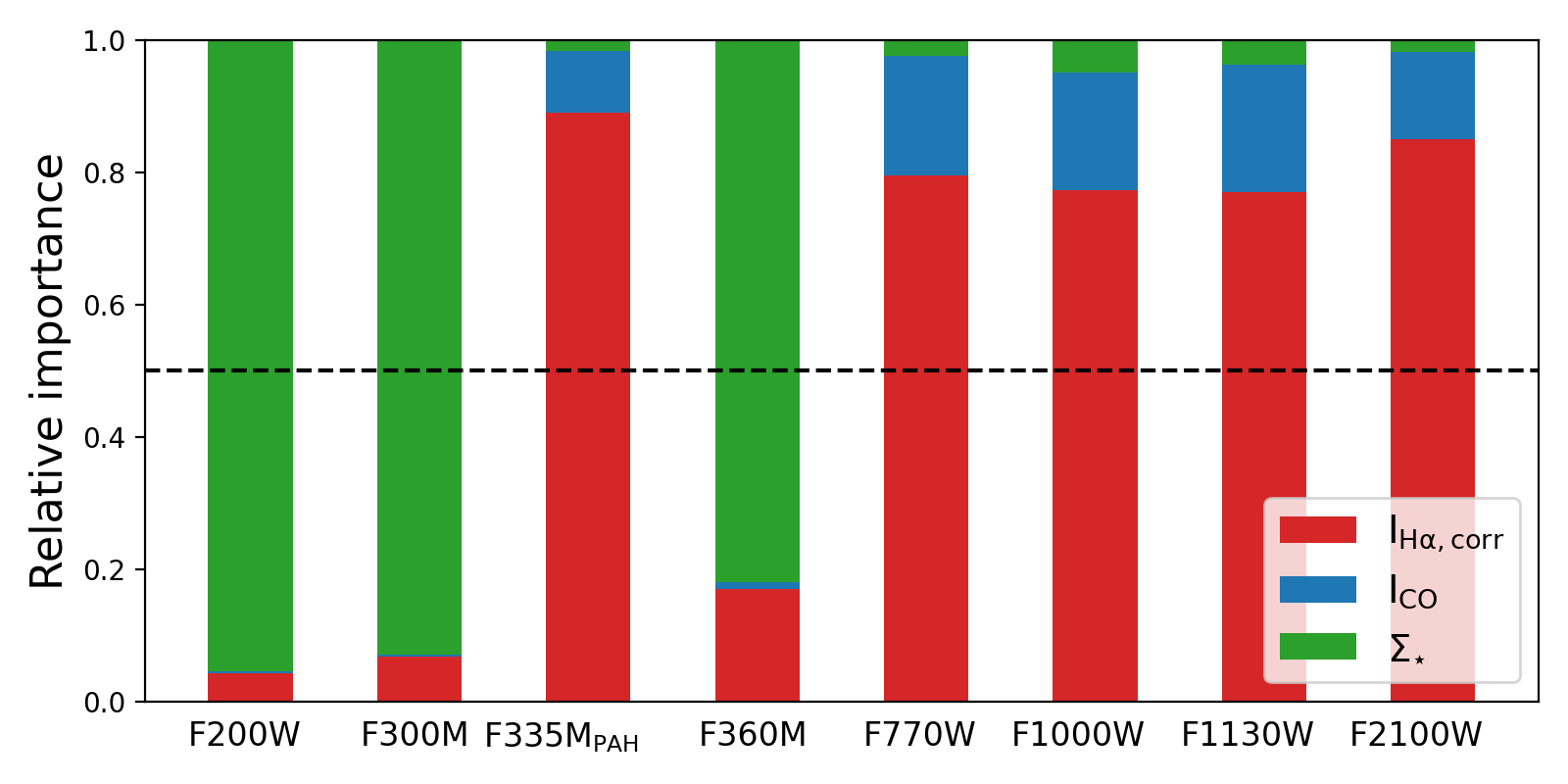}
\caption{Relative importance of extinction-corrected H$\alpha$ ($\rm I_{H\alpha, corr}$), CO(2-1) surface density, and stellar mass surface density ($\rm \Sigma_{\star}$) in predicting the surface brightness of \hii\ regions in a set of JWST bands. }
\label{fig:fig8}
\end{figure*}

\subsection{Does the IR emission trace embedded SFR or cold gas?}

The MIR, and in particular the bands containing strong PAH features, show excellent correlations with the cold molecular gas content, mapped by CO emission \citep{Regan2006, Chown2021, Whitcomb2023, Leroy2023a}. \cite{Leroy2021a} and \cite{Leroy2023a}, for example, found a nearly linear relation between the intensity in the WISE W3 band centred at 12~\um\ and CO(2-1) emission. This correlation is found to have lower scatter than that between CO and IR emission at $\sim$ 24 \um. \cite{Whitcomb2023} similarly argued using correlation analysis that emission around 12~\um\ traces CO emission better than SFR, while IR emission at 24~\um\ traces SFR better than CO. \cite{Leroy2023} use the early JWST data for four galaxies from our sample and compare the MIR emission with both \ha\ and CO emission, finding that MIR emission correlates well with both extinction-corrected \ha\ and CO. Such a result is expected, since dust emission depends on both the dust surface density and the intensity of the radiation field heating it. The convolution of these effects leads to MIR emission being a useful tracer of both $I_{\rm H\alpha, corr}$ and CO.

Within \hii\ regions, however, one expects a tighter relation between MIR emission and star formation than between MIR emission and cold gas content. To test this hypothesis we evaluate the relative ability of each of the JWST bands considered in this work to trace extinction-corrected \ha, molecular gas (as traced by CO(2-1)), and stellar mass surface density, focusing our analysis on \hii\ regions specifically. The analysis is performed by building a random forest regression model for the surface brightness in each JWST band and using $I_{\rm H\alpha, corr}$, $I_{\rm CO}$, and $\Sigma_{\star}$, in addition to  a vector of random numbers, as input features. The relative importance of each of the features in reproducing the brightness in each JWST band considered is presented in Fig. \ref{fig:fig8}. 

As expected, both F300M and F200W trace mostly stellar mass surface density (importance $\sim$ 90\%). For F360M \ha\ takes up part of the importance, probably reflecting the increasing importance of hot dust and residual contamination from PAHs in this band \citep{Sandstrom2023}.
F335M$_{\rm PAH}$ mostly traces \ha, highlighting the success of the continuum subtraction strategy for this band. The MIR bands trace mostly \ha, but with 20\% importance attributed also to CO for F770W, F1000W, F1130W. The contribution of stellar mass is small, and it is largest at F1000W, where is amounts to 5$\%$ of the overall importance. This band contains the largest contamination from point sources, potentially evolved stars, whose density correlates with the overall stellar mass surface density. For the canonical F2100W band the relative importance of \ha\ and CO are 83\% and 16\% respectively. In summary, both 3.3~\um\ and the MIR bands trace mostly extinction-corrected \ha\ within \hii\ regions, with a smaller ($<$20\%) secondary dependence on molecular gas, as traced by CO. The small dependence on CO is consistent with the finding that many \hii\ regions exists without associated molecular gas, in the sense of gas within the \hii\ region boundary \citep{Zakardjian2023}. These conclusions are likely to differ within the more diffuse medium, where the relative importance of CO in determining the MIR emission is expected to increase \citep{Leroy2023}, but we leave an exploration of these trends to future work.

\section{Conclusions}
\label{sec:conclusions}

We use NIRCam and MIRI images in combination with IFS from MUSE for 19 nearby galaxies to calibrate hybrid recipes to correct \ha\ for dust extinction on the scales of individual \hii\ regions ($\sim$ 100~pc). We examine hybrid recipes that consider a linear combination of \ha\ with the F2100W MIRI band, dominated by emission from small grains, and bands tracing PAH emission, namely the continuum-subtracted NIRCam F335M band (F335M$\rm _{PAH}$), and the F770W, F1000W, and F1130W MIRI bands. We summarise our main results below.
\begin{itemize}
    \item{The ratio between  21~\um\ and \ha\ emission is systematically higher in the diffuse ISM, outside optically defined \hii\ regions. This fact reflects a change in the dominant source of dust heating -- young stars in \hii\ regions and the old stellar population in the DIG -- and not an increase in dust attenuation. Fully embedded regions make a negligible contribution to the diffuse 21~\um\ emission.}
    \item{Focusing on \hii\ regions, we calibrate the coefficient in front of the IR emission term ($I_{\rm H\alpha, corr }= I_{\rm H\alpha} + a_{\rm band}~ I_{\rm band}$) by benchmarking the hybrid dust correction recipe with the correction obtained using the Balmer decrement measured from the MUSE data. We adopt fiducial hybridisation coefficients for the different dust-tracing JWST bands ($a_{\rm band}^{\rm C07}$) by rescaling the coefficient derived by \cite{Calzetti2007} for \textit{Spitzer} 24~\um\ emission by the average band ratio $\langle I_{24}/I_{\rm band} \rangle$. These C07-equivalent coefficients are in good general agreement with the median of the measured $a_{\rm band}$ coefficient across our sample of \hii\ regions for all bands considered (F335M$\rm _{PAH}$, F770W, F1000W, F1130W, F2100W). A local background subtraction causes an increase of the cofficients associated with F770W, F1000W, and F1130W bands of $\sim$ 0.4 dex, while it has a smaller effect on F2100W and F335M$_{\rm PAH}$ ($a$ increases by $\sim$ 0.2 dex). Our recommended coefficients are summarised in Table \ref{tab:recommend}.}
    \item{For both 21~\um\ and the PAH-tracing bands, using a hybrid recipe with a constant coefficient $a_{\rm band}^{\rm C07}$ leads to extinction-corrected \ha\ fluxes that correlate well (scatter $\sim$ 0.1 dex), albeit sub-linearly (slope 0.82-0.89), with those inferred from the Balmer decrement. The sub-linear slope implies that a constant $a_{\rm band}$ overestimates the dust correction for faint \hii\ regions, and it underestimates it for bright regions.}
    \item{We tested for secondary dependencies of $a_{\rm band}$ on a number of additional physical parameters. The strongest correlations are with physical parameters that trace the ratio between young and old stellar components, e.g. EW(\ha), sSFR, or indirectly correlate with them (e.g. \ha\ surface brightness, or dust attenuation measured from the Balmer decrement). Strong correlations are also found with line ratios that trace the change in the ionisation condition between \hii\ regions and DIG (\oi/\ha, \sii/\ha). These trends indicate that dust heating from the old stellar component contributes substantially to the IR emission for the fainter \hii\ regions in our sample.}
    \item{$a_{\rm band}$ correlates with EW(\ha) for all bands considered. For EW(\ha) $<$ 30 \AA, the trends are similar for all bands and agree with the recent kpc-resolution study of \cite{Belfiore2023}. At higher EW(\ha), $a_{21}$ plateaus to a constant value, while the coefficients for the F770W, F1000W, and F1130W keep increasing. The behaviour of the MIR PAH-tracing bands reflects the relative decrease in PAH emission in bright \hii\ regions, in agreement with the literature predicting the destruction of PAH molecules in such harsh environments. The F335M$\rm _{PAH}$ shows an intermediate behaviour between F2100W and the other MIR bands, which can be explained by the relative increase in 3.3~\um\ emission with respect to the other PAH-tracing bands in \hii\ regions, possibly tracing a decrease in the average size of PAH grains or an increase in the hardness of the UV radiation field.}
    \item{A random forest analysis confirms that F2100W is the preferred band to consider in combination with \ha\ to infer dust attenuation, confirming results from previous literature using WISE 22~\um\ or \textit{Spitzer} 24~\um\ bands.}
    \item{The main secondary dependence of $a_{\rm band}$ can be modelled by explicitly adding an additional linear term proportional to stellar mass surface density or 2~\um\ emission to the hybrid recipe. We provide coefficients for these corrections in Eq. \ref{eq:plane_fit2  } and \ref{eq:calib_star}. }
    \item{Finally, we analyse the relative importance of molecular gas (traced by CO(2-1)), stellar mass surface density, and extinction-corrected \ha\ in determining the surface brightness for \hii\ regions within our JWST bands. The MIR bands trace mostly \ha\, but with a smaller (15-20\%) importance attributed to CO.} 
\end{itemize}

\begin{acknowledgements}
This work is carried out as part of the PHANGS collaboration. 

Based on observations made with the NASA/ESA/CSA \textit{JWST}. The data were obtained from the Mikulski Archive for Space Telescopes at the Space Telescope Science Institute, which is operated by the Association of Universities for Research in Astronomy, Inc., under NASA contract NAS 5-03127. The observations are associated with \textit{JWST} program 2107 (PI: J. Lee)
Based on observations collected at the European Southern Observatory under ESO programmes 094.C-0623 (PI: Kreckel), 095.C-0473,  098.C-0484 (PI: Blanc), 1100.B-0651 (PHANGS-MUSE; PI: Schinnerer), as well as 094.B-0321 (MAGNUM; PI: Marconi), 099.B-0242, 0100.B-0116, 098.B-0551 (MAD; PI: Carollo) and 097.B-0640 (TIMER; PI: Gadotti).

FB acknowledges support from the INAF Fundamental Astrophysics program 2022.
MB acknowledges support from FONDECYT regular grant 1211000 and by the ANID BASAL project FB210003.

HAP acknowledges support by the National Science and Technology Council of Taiwan under grant 110-2112-M-032-020-MY3.

JMDK gratefully acknowledges funding from the ERC under the European Union's Horizon 2020 research and innovation programme via the ERC Starting Grant MUSTANG (grant number 714907). COOL Research DAO is a Decentralized Autonomous Organization supporting research in astrophysics aimed at uncovering our cosmic origins.

MC gratefully acknowledges funding from the DFG through an Emmy Noether Research Group (grant number CH2137/1-1).


RSK and SCOG acknowledge financial support from the ERC via the ERC Synergy Grant ``ECOGAL'' (project ID 855130), from the German Excellence Strategy via the Heidelberg Cluster of Excellence (EXC 2181 - 390900948) ``STRUCTURES'', and from the German Ministry for Economic Affairs and Climate Action in project ``MAINN'' (funding ID 50OO2206). RSK also thanks for computing resources provided bwHPC and DFG through grant INST 35/1134-1 FUGG and for data storage at SDS@hd through grant INST 35/1314-1 FUGG.
JC acknowledges support from ERC starting grant \#851622 DustOrigin.
MQ acknowledges support from the Spanish grant PID2019-106027GA-C44, funded by MCIN/AEI/10.13039/501100011033. 
KK, OVE and EJW gratefully acknowledge funding from the Deutsche Forschungsgemeinschaft (DFG, German Research Foundation) in the form of an Emmy Noether Research Group (grant number KR4598/2-1, PI Kreckel). 
JN acknowledges funding from the European Research Council (ERC) under the European Union’s Horizon 2020 research and innovation programme (grant agreement No. 694343).
ER and HH acknowledge the support of the Natural Sciences and Engineering Research Council of Canada (NSERC), funding reference number RGPIN-2022-03499, and the support of the Canadian Space Agency (22ASTALBER).
AKL gratefully acknowledges support by grants 1653300 and 2205628 from the National Science Foundation, by award JWST-GO-02107.009-A, and by a Humboldt Research Award from the Alexander von Humboldt Foundation. KS acknowledges funding support from grant support by JWST-GO-02107.006-A
\end{acknowledgements}


%
\bibliographystyle{aa} 
\bibliography{library7} 

\begin{thebibliography}{92}
\expandafter\ifx\csname natexlab\endcsname\relax\def\natexlab#1{#1}\fi

\bibitem[{Anand {et~al.}(2021)Anand, Lee, {Van Dyk}, Leroy, Rosolowsky,
  Schinnerer, Larson, Kourkchi, Kreckel, Scheuermann, Rizzi, Thilker, {Brent
  Tully}, Bigiel, Blanc, Boquien, Chandar, Dale, Emsellem, Deger, Glover,
  Grasha, Groves, Klessen, Kruijssen, Querejeta, S{\'{a}}nchez-Bl{\'{a}}zquez,
  Schruba, Turner, Ubeda, Williams, \& Whitmore}]{Anand2021}
Anand, G.~S., Lee, J.~C., {Van Dyk}, S.~D., {et~al.} 2021, MNRAS, 501, 3621

\bibitem[{Aniano {et~al.}(2011)Aniano, Draine, Gordon, \&
  Sandstrom}]{Aniano2011}
Aniano, G., Draine, B.~T., Gordon, K.~D., \& Sandstrom, K. 2011, PASP, 123,
  1218

\bibitem[{{Baker} {et~al.}(2022){Baker}, {Maiolino}, {Bluck}, {Lin}, {Ellison},
  {Belfiore}, {Pan}, \& {Thorp}}]{Baker2022}
{Baker}, W.~M., {Maiolino}, R., {Bluck}, A. F.~L., {et~al.} 2022, MNRAS, 510,
  3622

\bibitem[{Baldwin {et~al.}(1981)Baldwin, Phillips, \& Terlevich}]{Baldwin1981}
Baldwin, J.~A., Phillips, M.~M., \& Terlevich, R. 1981, PASP, 93, 5

\bibitem[{Belfiore {et~al.}(2023)Belfiore, Leroy, Sun, Barnes, Boquien, Cao,
  Congiu, Dale, Egorov, Eibensteiner, Glover, Grasha, Groves, Klessen, Kreckel,
  Neumann, Querejeta, Sanchez-Blazquez, Schinnerer, \& Williams}]{Belfiore2023}
Belfiore, F., Leroy, A., Sun, J., {et~al.} 2023, A{\&}A, 670, A67

\bibitem[{Belfiore {et~al.}(2022)Belfiore, Santoro, Groves, Schinnerer,
  Kreckel, Glover, Klessen, Emsellem, Blanc, Congiu, Barnes, Boquien, Chevance,
  Dale, {Diederik Kruijssen}, Leroy, Pan, Pessa, Schruba, \&
  Williams}]{Belfiore2022}
Belfiore, F., Santoro, F., Groves, B., {et~al.} 2022, A{\&}A, 659, A26

\bibitem[{Bendo {et~al.}(2012)Bendo, Boselli, Dariush, Pohlen, Roussel,
  Sauvage, Smith, Wilson, Baes, Cooray, Clements, Cortese, Foyle, Galametz,
  Gomez, Lebouteiller, Lu, Madden, Mentuch, O'Halloran, Page, Remy, Schulz, \&
  Spinoglio}]{Bendo2012}
Bendo, G.~J., Boselli, A., Dariush, A., {et~al.} 2012, MNRAS, 419, 1833

\bibitem[{Bendo {et~al.}(2006)Bendo, Dale, Draine, Engelbracht, {Kennicutt,
  Jr.}, Calzetti, Gordon, Helou, Hollenbach, Li, Murphy, Prescott, \&
  Smith}]{Bendo2006}
Bendo, G.~J., Dale, D.~A., Draine, B.~T., {et~al.} 2006, ApJ, 652, 283

\bibitem[{Bluck {et~al.}(2022)Bluck, Maiolino, Brownson, Conselice, Ellison,
  Piotrowska, \& Thorp}]{Bluck2022}
Bluck, A.~F., Maiolino, R., Brownson, S., {et~al.} 2022, A{\&}A, 659, A160

\bibitem[{Bolatto {et~al.}(2007)Bolatto, Simon, Stanimirovi{\'{c}}, van Loon,
  Shah, Venn, Leroy, Sandstrom, Jackson, Israel, Li, Staveley‐Smith, Bot,
  Boulanger, \& Rubio}]{Bolatto2007}
Bolatto, A.~D., Simon, J.~D., Stanimirovi{\'{c}}, S., {et~al.} 2007, ApJ, 655,
  212

\bibitem[{Bolatto {et~al.}(2013)Bolatto, Wolfire, \& Leroy}]{Bolatto2013}
Bolatto, A.~D., Wolfire, M., \& Leroy, A.~K. 2013, ARA{\&}A, 51, 207

\bibitem[{Boquien {et~al.}(2016)Boquien, Kennicutt, Calzetti, Dale, Galametz,
  Sauvage, Croxall, Draine, Kirkpatrick, Kumari, Hunt, {De Looze}, Pellegrini,
  Rela{\~{n}}o, Smith, \& Tabatabaei}]{Boquien2016}
Boquien, M., Kennicutt, R., Calzetti, D., {et~al.} 2016, A{\&}A, 591, A6

\bibitem[{Calzetti(2020)}]{Calzetti2020}
Calzetti, D. 2020, Nat. Astron., 4, 437

\bibitem[{Calzetti {et~al.}(2007)Calzetti, Kennicutt, Engelbracht, Leitherer,
  Draine, Kewley, Moustakas, Sosey, Dale, Gordon, Helou, Hollenbach, Armus,
  Bendo, Bot, Buckalew, Jarrett, Li, Meyer, Murphy, Prescott, Regan, Rieke,
  Roussel, Sheth, Smith, Thornley, \& Walter}]{Calzetti2007}
Calzetti, D., Kennicutt, R.~C., Engelbracht, C.~W., {et~al.} 2007, ApJ, 666,
  870

\bibitem[{Catal{\'{a}}n-Torrecilla {et~al.}(2015)Catal{\'{a}}n-Torrecilla, {Gil
  de Paz}, Castillo-Morales, Iglesias-P{\'{a}}ramo, S{\'{a}}nchez, Kennicutt,
  P{\'{e}}rez-Gonz{\'{a}}lez, Marino, Walcher, Husemann, Garc{\'{i}}a-Benito,
  Mast, {Gonz{\'{a}}lez Delgado}, Mu{\~{n}}oz-Mateos, Bland-Hawthorn, Bomans,
  {Del Olmo}, Galbany, Gomes, Kehrig, L{\'{o}}pez-S{\'{a}}nchez, Mendoza,
  Monreal-Ibero, P{\'{e}}rez-Torres, S{\'{a}}nchez-Bl{\'{a}}zquez, Vilchez, \&
  {Califa Collaboration}}]{Catalan-Torrecilla2015}
Catal{\'{a}}n-Torrecilla, C., {Gil de Paz}, A., Castillo-Morales, A., {et~al.}
  2015, A{\&}A, 584, A87

\bibitem[{Chastenet {et~al.}(2019)Chastenet, Sandstrom, Chiang, Leroy, Utomo,
  Bot, Gordon, Draine, Fukui, Onishi, \& Tsuge}]{Chastenet2019}
Chastenet, J., Sandstrom, K., Chiang, I.-D., {et~al.} 2019, ApJ, 876, 62

\bibitem[{Chastenet {et~al.}(2023{\natexlab{a}})Chastenet, Sutter, Sandstrom,
  Bel, Egorov, Larson, Leroy, Liu, Rosolowsky, Thilker, \&
  Watkins}]{Chastenet2023b}
Chastenet, J., Sutter, J., Sandstrom, K., {et~al.} 2023{\natexlab{a}}, ApJL,
  944, L12

\bibitem[{Chastenet {et~al.}(2023{\natexlab{b}})Chastenet, Sutter, Sandstrom,
  Belfiore, Egorov, Larson, Leroy, Liu, Rosolowsky, Thilker, Watkins, Williams,
  Barnes, Bigiel, Boquien, Chevance, Chiang, Dale, Kruijssen, Emsellem, Grasha,
  Groves, Hassani, Hughes, Kreckel, Meidt, Vaught, Sardone, \&
  Schinnerer}]{Chastenet2023a}
Chastenet, J., Sutter, J., Sandstrom, K., {et~al.} 2023{\natexlab{b}}, ApJL,
  944, L11

\bibitem[{Chown {et~al.}(2021)Chown, Li, Parker, Wilson, Li, \&
  Gao}]{Chown2021}
Chown, R., Li, C., Parker, L., {et~al.} 2021, MNRAS, 500, 1261

\bibitem[{Cortese {et~al.}(2008)Cortese, Boselli, Franzetti, Decarli, Gavazzi,
  Boissier, \& Buat}]{Cortese2008}
Cortese, L., Boselli, A., Franzetti, P., {et~al.} 2008, MNRAS, 386, 1157

\bibitem[{Crocker {et~al.}(2013)Crocker, Calzetti, Thilker, Aniano, Draine,
  Hunt, Kennicutt, Sandstrom, \& Smith}]{Crocker2013}
Crocker, A.~F., Calzetti, D., Thilker, D.~A., {et~al.} 2013, ApJ, 762, 79

\bibitem[{{De Looze} {et~al.}(2014){De Looze}, Fritz, Baes, Bendo, Cortese,
  Boquien, Boselli, Camps, Cooray, Cormier, Davies, {De Geyter}, Hughes, Jones,
  Karczewski, Lebouteiller, Lu, Madden, R{\'{e}}my-Ruyer, Spinoglio, Smith,
  Viaene, \& Wilson}]{DeLooze2014}
{De Looze}, I., Fritz, J., Baes, M., {et~al.} 2014, A{\&}A, 571, 1

\bibitem[{{Den Brok} {et~al.}(2021){Den Brok}, Chatzigiannakis, Bigiel,
  Puschnig, Barnes, Leroy, Jim{\'{e}}nez-Donaire, Usero, Schinnerer,
  Rosolowsky, Faesi, Grasha, Hughes, Kruijssen, Liu, Neumann, Pety, Querejeta,
  Saito, Schruba, \& Stuber}]{DenBrok2021}
{Den Brok}, J.~S., Chatzigiannakis, D., Bigiel, F., {et~al.} 2021, MNRAS, 504,
  3221

\bibitem[{Diaz {et~al.}(1991)Diaz, Terlevich, Vflchez, Pagel, Edmunds,
  Te{\'{o}}rica, \& Madrid}]{Diaz1991}
Diaz, A.~I., Terlevich, E., Vflchez, J.~M., {et~al.} 1991, MNRAS, 253, 245

\bibitem[{Draine \& Li(2007)}]{Draine2007}
Draine, B.~T. \& Li, A. 2007, ApJ, 657, 810

\bibitem[{Draine {et~al.}(2021)Draine, Li, Hensley, Hunt, Sandstrom, \&
  Smith}]{Draine2021}
Draine, B.~T., Li, A., Hensley, B.~S., {et~al.} 2021, ApJ, 917, 3

\bibitem[{{Egorov} {et~al.}(2023){Egorov}, {Kreckel}, {Sandstrom}, {Leroy},
  {Glover}, {Groves}, {Kruijssen}, {Barnes}, {Belfiore}, {Bigiel}, {Blanc},
  {Boquien}, {Cao}, {Chastenet}, {Chevance}, {Congiu}, {Dale}, {Emsellem},
  {Grasha}, {Klessen}, {Larson}, {Liu}, {Murphy}, {Pan}, {Pessa}, {Pety},
  {Rosolowsky}, {Scheuermann}, {Schinnerer}, {Sutter}, {Thilker}, {Watkins}, \&
  {Williams}}]{Egorov2023}
{Egorov}, O.~V., {Kreckel}, K., {Sandstrom}, K.~M., {et~al.} 2023, ApJL, 944,
  L16

\bibitem[{Emsellem {et~al.}(2022)Emsellem, Schinnerer, Santoro, Belfiore,
  Pessa, McElroy, Blanc, Congiu, Groves, Ho, Kreckel, Razza, Sanchez-Blazquez,
  Egorov, Faesi, Klessen, Leroy, Meidt, Querejeta, Rosolowsky, Scheuermann,
  Anand, Barnes, Bes, Bigiel, Boquien, Cao, Chevance, Dale, Eibensteiner,
  Glover, Grasha, Henshaw, Hughes, Koch, Kruijssen, Lee, Liu, Pan, Pety, Saito,
  Sandstrom, Schruba, Sun, Thilker, Usero, Watkins, \& Williams}]{Emsellem2022}
Emsellem, E., Schinnerer, E., Santoro, F., {et~al.} 2022, A{\&}A, 659, A191

\bibitem[{{Galliano} {et~al.}(2018){Galliano}, {Galametz}, \&
  {Jones}}]{Galliano2018}
{Galliano}, F., {Galametz}, M., \& {Jones}, A.~P. 2018, ARA\&A, 56, 673

\bibitem[{Groves {et~al.}(2012)Groves, Krause, Sandstrom, Schmiedeke, Leroy,
  Linz, Kapala, Rix, Schinnerer, Tabatabaei, Walter, \& da~Cunha}]{Groves2012}
Groves, B., Krause, O., Sandstrom, K., {et~al.} 2012, MNRAS, 426, 892

\bibitem[{Groves {et~al.}(2023)Groves, Kreckel, Santoro, Belfiore, Zavodnik,
  Congiu, Egorov, Emsellem, Grasha, Leroy, Scheuermann, Schinnerer, Watkins,
  Barnes, Bigiel, Dale, Glover, Pessa, Sanchez-Blazquez, \&
  Williams}]{Groves2023}
Groves, B., Kreckel, K., Santoro, F., {et~al.} 2023, MNRAS, 520, 4902

\bibitem[{Gruppioni {et~al.}(2013)Gruppioni, Pozzi, Rodighiero, Delvecchio,
  Berta, Pozzetti, Zamorani, Andreani, Cimatti, Ilbert, {Le Floc'h}, Lutz,
  Magnelli, Marchetti, Monaco, Nordon, Oliver, Popesso, Riguccini, Roseboom,
  Rosario, Sargent, Vaccari, Altieri, Aussel, Bongiovanni, Cepa, Daddi,
  Dom{\'{i}}nguez-S{\'{a}}nchez, Elbaz, Schreiber, Genzel, Iribarrem,
  Magliocchetti, Maiolino, Poglitsch, Garc{\'{i}}a, Sanchez-Portal, Sturm,
  Tacconi, Valtchanov, Amblard, Arumugam, Bethermin, Bock, Boselli, Buat,
  Burgarella, Castro-Rodr{\'{i}}guez, Cava, Chanial, Clements, Conley, Cooray,
  Dowell, Dwek, Eales, Franceschini, Glenn, Griffin, Hatziminaoglou, Ibar,
  Isaak, Ivison, Lagache, Levenson, Lu, Madden, Maffei, Mainetti, Nguyen,
  O'Halloran, Page, Panuzzo, Papageorgiou, Pearson, P{\'{e}}rez-Fournon,
  Pohlen, Rigopoulou, Rowan-Robinson, Schulz, Scott, Seymour, Shupe, Smith,
  Stevens, Symeonidis, Trichas, Tugwell, Vigroux, Wang, Wright, Xu, Zemcov,
  Bardelli, Carollo, Contini, {Le F{\'{e}}vre}, Lilly, Mainieri, Renzini,
  Scodeggio, \& Zucca}]{Gruppioni2013}
Gruppioni, C., Pozzi, F., Rodighiero, G., {et~al.} 2013, MNRAS, 432, 23

\bibitem[{Haffner {et~al.}(2009)Haffner, Dettmar, Beckman, Wood, Slavin,
  Giammanco, Madsen, Zurita, \& Reynolds}]{Haffner2009}
Haffner, L.~M., Dettmar, R.-J.~J., Beckman, J.~E., {et~al.} 2009, Rev. Mod.
  Phys., 81, 969

\bibitem[{Hao {et~al.}(2011)Hao, Kennicutt, Johnson, Calzetti, Dale, \&
  Moustakas}]{Hao2011}
Hao, C.~N., Kennicutt, R.~C., Johnson, B.~D., {et~al.} 2011, ApJ, 741, 124

\bibitem[{Hassani {et~al.}(2023)Hassani, Rosolowsky, Leroy, Boquien, Lee,
  Barnes, Bel, Bigiel, Cao, Chevance, Dale, \& Egorov}]{Hassani2023}
Hassani, H., Rosolowsky, E., Leroy, A.~K., {et~al.} 2023, ApJL, 944, L21

\bibitem[{Helou {et~al.}(2004)Helou, Roussel, Appleton, Frayer, Stolovy,
  Storrie‐Lombardi, Hurt, Lowrance, Makovoz, Masci, Surace, Gordon,
  Alonso‐Herrero, Engelbracht, Misselt, Rieke, Rieke, Willner, Pahre, Ashby,
  Fazio, \& Smith}]{Helou2004}
Helou, G., Roussel, H., Appleton, P., {et~al.} 2004, ApJS, 154, 253

\bibitem[{Hirashita {et~al.}(2003)Hirashita, Buat, \& Inoue}]{Hirashita2003}
Hirashita, H., Buat, V., \& Inoue, A.~K. 2003, A{\&}A, 410, 83

\bibitem[{Hogg {et~al.}(2010)Hogg, Bovy, \& Lang}]{Hogg2010a}
Hogg, D.~W., Bovy, J., \& Lang, D. 2010 [\eprint[arXiv]{1008.4686}]

\bibitem[{Jarrett {et~al.}(2013)Jarrett, Masci, Tsai, Petty, Cluver, Assef,
  Benford, Blain, Bridge, Donoso, Eisenhardt, Koribalski, Lake, Neill, Seibert,
  Sheth, Stanford, \& Wright}]{Jarrett2013}
Jarrett, T.~H., Masci, F., Tsai, C.~W., {et~al.} 2013, AJ, 145, 6

\bibitem[{Kauffmann {et~al.}(2003)Kauffmann, Heckman, Tremonti, Brinchmann,
  Charlot, White, Ridgway, Brinkmann, Fukugita, Hall, Ivezi{\'{c}}, Richards,
  \& Schneider}]{Kauffmann2003a}
Kauffmann, G., Heckman, T.~M., Tremonti, C., {et~al.} 2003, MNRAS, 346, 1055

\bibitem[{Kelly(2007)}]{Kelly2007}
Kelly, B.~C. 2007, ApJ, 665, 1489

\bibitem[{Kennicutt(1998)}]{Kennicutt1998}
Kennicutt, R.~C. 1998, ARA{\&}A, 36, 189

\bibitem[{Kennicutt {et~al.}(2007)Kennicutt, Calzetti, Walter, Helou,
  Hollenbach, Armus, Bendo, Dale, Draine, Engelbracht, Gordon, Prescott, Regan,
  Thornley, Bot, Brinks, de~Blok, de~Mello, Meyer, Moustakas, Murphy, Sheth, \&
  Smith}]{Kennicutt2007}
Kennicutt, R.~C., Calzetti, D., Walter, F., {et~al.} 2007, ApJ, 671, 333

\bibitem[{Kennicutt \& Evans(2012)}]{Kennicutt2012}
Kennicutt, R.~C. \& Evans, N.~J. 2012, ARA{\&}A, 50, 531

\bibitem[{Kennicutt {et~al.}(2009)Kennicutt, Hao, Calzetti, Moustakas, Dale,
  Bendo, Engelbracht, Johnson, \& Lee}]{Kennicutt2009}
Kennicutt, R.~C., Hao, C.~N., Calzetti, D., {et~al.} 2009, ApJ, 703, 1672

\bibitem[{Kewley {et~al.}(2001)Kewley, Dopita, Sutherland, Heisler, \&
  Trevena}]{Kewley2001}
Kewley, L.~J., Dopita, M.~A., Sutherland, R.~S., Heisler, C.~A., \& Trevena, J.
  2001, ApJ, 556, 121

\bibitem[{{Khramtsova} {et~al.}(2013){Khramtsova}, {Wiebe}, {Boley}, \&
  {Pavlyuchenkov}}]{Khramtsova2013}
{Khramtsova}, M.~S., {Wiebe}, D.~S., {Boley}, P.~A., \& {Pavlyuchenkov}, Y.~N.
  2013, MNRAS, 431, 2006

\bibitem[{Kim {et~al.}(2022)Kim, Chevance, Kruijssen, Leroy, Schruba, Barnes,
  Bigiel, Blanc, Cao, Congiu, Dale, Faesi, Glover, Grasha, Groves, Hughes,
  Klessen, Kreckel, McElroy, Pan, Pety, Querejeta, Razza, Rosolowsky, Saito,
  Schinnerer, Sun, Tomi{\v{c}}i{\'{c}}, Usero, \& Williams}]{Kim2022a}
Kim, J., Chevance, M., Kruijssen, J. M.~D., {et~al.} 2022, MNRAS, 516, 3006

\bibitem[{{Kouroumpatzakis} {et~al.}(2023){Kouroumpatzakis}, {Zezas},
  {Kyritsis}, {Salim}, \& {Svoboda}}]{Kouroumpatzakis2023}
{Kouroumpatzakis}, K., {Zezas}, A., {Kyritsis}, E., {Salim}, S., \& {Svoboda},
  J. 2023, A\&A, 673, A16

\bibitem[{Lai {et~al.}(2020)Lai, Smith, Baba, Spoon, \& Imanishi}]{Lai2020a}
Lai, T. S.-Y., Smith, J. D.~T., Baba, S., Spoon, H. W.~W., \& Imanishi, M.
  2020, ApJ, 905, 55

\bibitem[{Lang {et~al.}(2020)Lang, Meidt, Rosolowsky, Nofech, Schinnerer,
  Leroy, Emsellem, Pessa, Glover, Groves, Hughes, Kruijssen, Querejeta,
  Schruba, Bigiel, Blanc, Chevance, Colombo, Faesi, Henshaw, Herrera, Liu,
  Pety, Puschnig, Saito, Sun, \& Usero}]{Lang2020}
Lang, P., Meidt, S.~E., Rosolowsky, E., {et~al.} 2020, ApJ, 897, 122

\bibitem[{Lebouteiller {et~al.}(2011)Lebouteiller, Bernard-Salas, Whelan,
  Brandl, Galliano, Charmandaris, Madden, \& Kunth}]{Lebouteiller2011}
Lebouteiller, V., Bernard-Salas, J., Whelan, D.~G., {et~al.} 2011, ApJ, 728

\bibitem[{Lee {et~al.}(2023)Lee, Sandstrom, Leroy, Thilker, Schinnerer,
  Rosolowsky, Larson, Egorov, Williams, Schmidt, Emsellem, Anand, Barnes,
  Belfiore, Beslic, Bigiel, Blanc, Bolatto, Boquien, den Brok, Cao, Chandar,
  Chastenet, Chevance, Chiang, Congiu, Dale, Deger, Eibensteiner, Faesi,
  Glover, Grasha, Groves, Hassani, Henny, Henshaw, Hoyer, Hughes, Jeffreson,
  Jimenez-Donaire, Kim, Kim, Klessen, Koch, Kreckel, Kruijssen, Li, Liu, Lopez,
  Maschmann, Chen, Meidt, Murphy, Neumann, Neumayer, Pan, Pessa, Pety,
  Querejeta, Pinna, Rodrıguez, Saito, Sanchez-Blazquez, Santoro, Sardone,
  Smith, Sormani, Scheuermann, Stuber, Sutter, Sun, Teng, Tress, Usero,
  Watkins, Whitmore, \& Razza}]{Lee2023}
Lee, J.~C., Sandstrom, K.~M., Leroy, A.~K., {et~al.} 2023, ApJL, 944, L17

\bibitem[{Leroy {et~al.}(2012)Leroy, Bigiel, de~Blok, Boissier, Bolatto,
  Brinks, Madore, Munoz-Mateos, Murphy, Sandstrom, Schruba, \&
  Walter}]{Leroy2012}
Leroy, A.~K., Bigiel, F., de~Blok, W. J.~G., {et~al.} 2012, AJ, 144, 3

\bibitem[{Leroy {et~al.}(2023{\natexlab{a}})Leroy, Bolatto, Sandstrom,
  Rosolowsky, Barnes, Bigiel, Boquien, den Brok, Cao, Chastenet, Chevance,
  Chiang, Chown, Colombo, Ellison, Emsellem, Grasha, Henshaw, Hughes, Klessen,
  Koch, Kim, Kreckel, Kruijssen, Larson, Lee, Levy, Lin, Liu, Meidt, Pety,
  Querejeta, Rubio, Saito, Salim, Schinnerer, Sormani, Sun, Thilker, Usero,
  Vogel, Watkins, Whitcomb, Williams, \& Wilson}]{Leroy2023a}
Leroy, A.~K., Bolatto, A.~D., Sandstrom, K., {et~al.} 2023{\natexlab{a}}, ApJ,
  944, L10

\bibitem[{Leroy {et~al.}(2021{\natexlab{a}})Leroy, Hughes, Liu, Pety,
  Rosolowsky, Saito, Schinnerer, Schruba, Usero, Faesi, Herrera, Chevance,
  Hygate, Kepley, Koch, Querejeta, Sliwa, Will, Wilson, Anand, Barnes,
  Belfiore, Be{\v{s}}li{\'{c}}, Bigiel, Blanc, Bolatto, Boquien, Cao, Chandar,
  Chastenet, Chiang, Congiu, Dale, Deger, den Brok, Eibensteiner, Emsellem,
  Garc{\'{i}}a-Rodr{\'{i}}guez, Glover, Grasha, Groves, Henshaw, {Jim{\'{e}}nez
  Donaire}, Kim, Klessen, Kreckel, Kruijssen, Larson, Lee, Mayker, McElroy,
  Meidt, Mok, Pan, Puschnig, Razza, S{\'{a}}nchez-Bl'azquez, Sandstrom,
  Santoro, Sardone, Scheuermann, Sun, Thilker, Turner, Ubeda, Utomo, Watkins,
  \& Williams}]{Leroy2021b}
Leroy, A.~K., Hughes, A., Liu, D., {et~al.} 2021{\natexlab{a}}, ApJS, 255, 19

\bibitem[{Leroy {et~al.}(2023{\natexlab{b}})Leroy, Sandstrom, Rosolowsky, Bel,
  Bolatto, \& Cao}]{Leroy2023}
Leroy, A.~K., Sandstrom, K., Rosolowsky, E., {et~al.} 2023{\natexlab{b}}, ApJL,
  944, L9

\bibitem[{Leroy {et~al.}(2019)Leroy, Sandstrom, Lang, Lewis, Salim, Behrens,
  Chastenet, Chiang, Gallagher, Kessler, \& Utomo}]{Leroy2019}
Leroy, A.~K., Sandstrom, K.~M., Lang, D., {et~al.} 2019, ApJS, 244, 24

\bibitem[{Leroy {et~al.}(2021{\natexlab{b}})Leroy, Schinnerer, Hughes,
  Rosolowsky, Pety, Schruba, Usero, Blanc, Chevance, Emsellem, Faesi, Herrera,
  Liu, Meidt, Querejeta, Saito, Sandstrom, Sun, Williams, Anand, Barnes,
  Behrens, Belfiore, Benincasa, Be{\v{s}}li{\'{c}}, Bigiel, Bolatto, den Brok,
  Cao, Chandar, Chastenet, Chiang, Congiu, Dale, Deger, Eibensteiner, Egorov,
  Garc{\'{i}}a-Rodr{\'{i}}guez, Glover, Grasha, Henshaw, Ho, Kepley, Kim,
  Klessen, Kreckel, Koch, Kruijssen, Larson, Lee, Lopez, Machado, Mayker,
  McElroy, Murphy, Ostriker, Pan, Pessa, Puschnig, Razza,
  S{\'{a}}nchez-Bl{\'{a}}zquez, Santoro, Sardone, Scheuermann, Sliwa, Sormani,
  Stuber, Thilker, Turner, Utomo, Watkins, \& Whitmore}]{Leroy2021a}
Leroy, A.~K., Schinnerer, E., Hughes, A., {et~al.} 2021{\natexlab{b}}, ApJS,
  257, 43

\bibitem[{Leslie {et~al.}(2018)Leslie, Schinnerer, Groves, Sargent, Zamorani,
  Lang, \& Vardoulaki}]{Lesie2018}
Leslie, S.~K., Schinnerer, E., Groves, B., {et~al.} 2018, A{\&}A, 616, 1

\bibitem[{Li(2020)}]{Li2020}
Li, A. 2020, Nat. Astron., 4, 339

\bibitem[{Lu {et~al.}(2015)Lu, Blanc, \& Benson}]{Lu2015}
Lu, Y., Blanc, G.~A., \& Benson, A. 2015, ApJ, 808, 129

\bibitem[{Lutz {et~al.}(2005)Lutz, Valiante, Sturm, Genzel, Tacconi, Lehnert,
  Sternberg, \& Baker}]{Lutz2005}
Lutz, D., Valiante, E., Sturm, E., {et~al.} 2005, ApJ, 625, L83

\bibitem[{Madau \& Dickinson(2014)}]{Madau2014}
Madau, P. \& Dickinson, M. 2014, ARA{\&}A, 52, 415

\bibitem[{{Madden} {et~al.}(2006){Madden}, {Galliano}, {Jones}, \&
  {Sauvage}}]{Madden2006}
{Madden}, S.~C., {Galliano}, F., {Jones}, A.~P., \& {Sauvage}, M. 2006, A\&A,
  446, 877

\bibitem[{Maragkoudakis {et~al.}(2018)Maragkoudakis, Ivkovich, Peeters, Stock,
  Hemachandra, \& Tielens}]{Maragkoudakis2018}
Maragkoudakis, A., Ivkovich, N., Peeters, E., {et~al.} 2018, MNRAS, 481, 5370

\bibitem[{Nersesian {et~al.}(2019)Nersesian, Xilouris, Bianchi, Galliano,
  Jones, Baes, Casasola, Cassar{\`{a}}, Clark, Davies, Decleir, Dobbels, {De
  Looze}, {De Vis}, Fritz, Galametz, Madden, Mosenkov, Tr{\v{c}}ka, Verstocken,
  Viaene, \& Lianou}]{Nersesian2019}
Nersesian, A., Xilouris, E.~M., Bianchi, S., {et~al.} 2019, A{\&}A, 624, A80

\bibitem[{O'Donnell(1994)}]{O'Donnell1994}
O'Donnell, J.~E. 1994, ApJ, 422, 158

\bibitem[{Osterbrock \& Ferland(2006)}]{Osterbrock2006}
Osterbrock, D.~E. \& Ferland, G.~J. 2006, {Astrophysics of Gaseous Nebulae and
  Active Galactic Nuclei.}

\bibitem[{Pedregosa {et~al.}(2011)Pedregosa, Varoquaux, Gramfort, Michel,
  Thirion, Grisel, Blondel, M{\"{u}}ller, Nothman, Louppe, Prettenhofer, Weiss,
  Dubourg, Vanderplas, Passos, Cournapeau, Brucher, Perrot, \&
  Duchesnay}]{Pedregosa2011}
Pedregosa, F., Varoquaux, G., Gramfort, A., {et~al.} 2011, JMLR, 12, 2825

\bibitem[{Peeters {et~al.}(2004)Peeters, Spoon, \& Tielens}]{Peeters2004}
Peeters, E., Spoon, H. W.~W., \& Tielens, A. G. G.~M. 2004, ApJ, 613, 986

\bibitem[{Perrin {et~al.}(2014)Perrin, Sivaramakrishnan, Lajoie, Elliott,
  Pueyo, Ravindranath, \& Albert}]{Perrin2014}
Perrin, M.~D., Sivaramakrishnan, A., Lajoie, C.-P., {et~al.} 2014, Sp. Telesc.
  Instrum. 2014 Opt. Infrared, Millim. Wave, 9143, 91433X

\bibitem[{Phillips {et~al.}(1986)Phillips, Jenkins, Dopita, Sadler, \&
  Binette}]{Phillips1986}
Phillips, M.~M., Jenkins, E.~B., Dopita, M.~A., Sadler, E.~M., \& Binette, L.
  1986, AJ, 91, 1062

\bibitem[{Pilyugin \& Grebel(2016)}]{Pilyugin2016}
Pilyugin, L.~S. \& Grebel, E.~K. 2016, MNRAS, 457, 3678

\bibitem[{Regan {et~al.}(2006)Regan, Thornley, Vogel, Sheth, Draine,
  Hollenbach, Meyer, Dale, Engelbracht, Kennicutt, Armus, Buckalew, Calzetti,
  Gordon, Helou, Leitherer, Malhotra, Murphy, Rieke, Rieke, \&
  Smith}]{Regan2006}
Regan, M.~W., Thornley, M.~D., Vogel, S.~N., {et~al.} 2006, ApJ, 652, 1112

\bibitem[{Riechers {et~al.}(2014)Riechers, Pope, Daddi, Armus, Carilli, Walter,
  Hodge, Chary, Morrison, Dickinson, Dannerbauer, \& Elbaz}]{Riechers2014}
Riechers, D.~A., Pope, A., Daddi, E., {et~al.} 2014, ApJ, 786, 31

\bibitem[{Salim {et~al.}(2018)Salim, Boquien, \& Lee}]{Salim2018}
Salim, S., Boquien, M., \& Lee, J.~C. 2018, ApJ, 859, 11

\bibitem[{Sandstrom {et~al.}(2023{\natexlab{a}})Sandstrom, Chastenet, Sutter,
  Leroy, Egorov, Williams, Bolatto, Boquien, Cao, Dale, Lee, Rosolowsky,
  Schinnerer, Barnes, Belfiore, Bigiel, Chevance, Grasha, Groves, Hassani,
  Hughes, Klessen, Kruijssen, Larson, Liu, Lopez, Meidt, Murphy, Sormani,
  Thilker, \& Watkins}]{Sandstrom2023}
Sandstrom, K.~M., Chastenet, J., Sutter, J., {et~al.} 2023{\natexlab{a}}, ApJ,
  944, L7

\bibitem[{Sandstrom {et~al.}(2023{\natexlab{b}})Sandstrom, Koch, Leroy,
  Rosolowsky, Emsellem, Smith, Egorov, Williams, Larson, Lee, \&
  Schinnerer}]{Sandstrom2023a}
Sandstrom, K.~M., Koch, E.~W., Leroy, A.~K., {et~al.} 2023{\natexlab{b}}, ApJL,
  944, L8

\bibitem[{Shipley {et~al.}(2016)Shipley, Papovich, Rieke, Brown, \&
  Moustakas}]{Shipley2016a}
Shipley, H.~V., Papovich, C., Rieke, G.~H., Brown, M. J.~I., \& Moustakas, J.
  2016, ApJ, 818, 60

\bibitem[{Smith {et~al.}(2007)Smith, Draine, Dale, Moustakas, {Kennicutt, Jr.},
  Helou, Armus, Roussel, Sheth, Bendo, Buckalew, Calzetti, Engelbracht, Gordon,
  Hollenbach, Li, Malhotra, Murphy, \& Walter}]{Smith2007}
Smith, J. D.~T., Draine, B.~T., Dale, D.~A., {et~al.} 2007, ApJ, 656, 770

\bibitem[{Thilker {et~al.}(2000)Thilker, Braun, \& Walterbos}]{Thilker2000}
Thilker, D.~A., Braun, R., \& Walterbos, R. A.~M. 2000, AJ, 120, 3070

\bibitem[{Viaene {et~al.}(2017)Viaene, Sarzi, Baes, Fritz, \&
  Puerari}]{Viaene2017}
Viaene, S., Sarzi, M., Baes, M., Fritz, J., \& Puerari, I. 2017, MNRAS, 472,
  1286

\bibitem[{Westfall {et~al.}(2019)Westfall, Cappellari, Bershady, Bundy,
  Belfiore, Ji, Law, Schaefer, Shetty, Tremonti, Yan, Andrews, Brownstein,
  Cherinka, Coccato, Drory, Maraston, Parikh, Sanchez-Gallego, Thomas,
  Weijmans, Barrera-Ballesteros, Du, Goddard, Li, Masters, Medel, Sanchez,
  Yang, Zheng, \& Zhou}]{Westfall2019}
Westfall, K.~B., Cappellari, M., Bershady, M.~A., {et~al.} 2019, AJ, 158, 57

\bibitem[{{Whitcomb} {et~al.}(2023){Whitcomb}, {Sandstrom}, {Leroy}, \&
  {Smith}}]{Whitcomb2023}
{Whitcomb}, C.~M., {Sandstrom}, K., {Leroy}, A., \& {Smith}, J. D.~T. 2023,
  ApJ, 948, 88

\bibitem[{Whitcomb {et~al.}(2020)Whitcomb, Sandstrom, Murphy, \&
  Linden}]{Whitcomb2020}
Whitcomb, C.~M., Sandstrom, K., Murphy, E.~J., \& Linden, S. 2020, ApJ, 901, 47

\bibitem[{Williams {et~al.}(2019)Williams, Baes, Looze, Rela{\~{n}}o, Smith,
  Verstocken, \& Viaene}]{Williams2019}
Williams, T.~G., Baes, M., Looze, I.~D., {et~al.} 2019, MNRAS, 487, 2753

\bibitem[{Wuyts {et~al.}(2011{\natexlab{a}})Wuyts, {Forster Schreiber}, Lutz,
  Nordon, Berta, Altieri, Bongiovanni, Cepa, Cimatti, Daddi, Andreani, Elbaz,
  Genzel, Koekemoer, Magnelli, Maiolino, Poglitsch, Popesso, Pozzi, Mcgrath,
  Ana, Sanchez-portal, Sturm, Tacconi, \& Valtchanov}]{Wuyts2011a}
Wuyts, S., {Forster Schreiber}, N.~M., Lutz, D., {et~al.} 2011{\natexlab{a}},
  ApJ, 738, 106

\bibitem[{Wuyts {et~al.}(2011{\natexlab{b}})Wuyts, {F{\"{o}}rster Schreiber},
  van~der Wel, Magnelli, Guo, Genzel, Lutz, Aussel, Barro, Berta, Cava,
  Graci{\'{a}}-Carpio, Hathi, Huang, Kocevski, Koekemoer, Lee, {Le Floc'h},
  McGrath, Nordon, Popesso, Pozzi, Riguccini, Rodighiero, Saintonge, \&
  Tacconi}]{Wuyts2011}
Wuyts, S., {F{\"{o}}rster Schreiber}, N.~M., van~der Wel, A., {et~al.}
  2011{\natexlab{b}}, ApJ, 742, 96

\bibitem[{{Zakardjian} {et~al.}(2023){Zakardjian}, {Pety}, {Herrera}, {Hughes},
  {Oakes}, {Kreckel}, {Faesi}, {Glover}, {Groves}, {Klessen}, {Meidt},
  {Barnes}, {Belfiore}, {Be{\v{s}}li{\'c}}, {Bigiel}, {Blanc}, {Chevance},
  {Dale}, {den Brok}, {Eibensteiner}, {Emsellem}, {Garc{\'\i}a-Rodr{\'\i}guez},
  {Grasha}, {Koch}, {Leroy}, {Liu}, {Mc Elroy}, {Neumann}, {Pan}, {Querejeta},
  {Razza}, {Rosolowsky}, {Saito}, {Santoro}, {Schinnerer}, {Sun}, {Usero},
  {Watkins}, \& {Williams}}]{Zakardjian2023}
{Zakardjian}, A., {Pety}, J., {Herrera}, C.~N., {et~al.} 2023, arXiv,
  arXiv:2305.03650

\bibitem[{Zhang {et~al.}(2017)Zhang, Yan, Bundy, Bershady, Haffner, Walterbos,
  Maiolino, Tremonti, Thomas, Drory, Jones, Belfiore, S{\'{a}}nchez,
  Diamond-stanic, Bizyaev, Nitschelm, Andrews, Brinkmann, Brownstein, Cheung,
  Li, Law, {Roman Lopes}, Oravetz, Pan, {Storchi Bergmann}, Simmons, Sebastian,
  Diamond-stanic, Bizyaev, Nitschelm, Andrews, Brinkmann, Brownstein, Cheung,
  Li, Law, Lopes, \& Oravetz}]{Zhang2017}
Zhang, K., Yan, R., Bundy, K., {et~al.} 2017, MNRAS, 466, 3217

\bibitem[{Zhang \& Ho(2023)}]{Zhang2023}
Zhang, L. \& Ho, L.~C. 2023, ApJ, 943, 60

\end{thebibliography}




\end{document}